\documentclass[epj]{svjour}
\usepackage{graphics,amsmath}

\def\gs{\gamma_S}
\def\Nj{\{N_j\}}
\def\hpart{\{h_n\}}
\def\hpartj{\{h_{n_j}\}}
\def\Qz{{\bf Q}}
\def\Qs{{\bf Q}_{\Nj}}
\def\qj{{\bf q}_j}
\def\phiv{\mbox{\boldmath$\phi$}}
\def\phivs{\mbox{\boldmath${\scriptstyle{\phi}}$}}
\def\pivs{\mbox{\boldmath${\scriptstyle{\pi}}$}}

\def\d{{\rm d}}

\def\e{{\rm e}}
\def\i{{\rm i}}

\def\inty{\int_{-\infty-\i \varepsilon}^{+\infty-\i \varepsilon} 
\!\!\!\!\!\!\!\! \d^4 y}
\def\intyy{\int_{-\infty-\i \varepsilon}^{+\infty-\i \varepsilon} 
\!\!\!\!\!\!\!\!\!\!\!\!\!\! \d^4 y}
\def\intz{\int_{-\infty-\i \varepsilon}^{+\infty-\i \varepsilon} 
\!\!\!\!\!\!\!\! \d^4 z}

\def\ssb{\langle {\rm s}\bar{\rm s}\rangle}

\def\ppb{${\rm p}\bar{\rm p}\;$}
\begin{document}
\title{Statistical hadronization and hadronic microcanonical ensemble I}
\author{F. Becattini\inst{1} \and L. Ferroni\inst{1}
}                     
\institute{Universit\`a di Firenze and INFN Sezione di Firenze}
%
%
\abstract{
We present a full treatment of the microcanonical ensemble of the ideal 
hadron-resonance gas in a quantum-mechanical framework which is appropriate 
for the statistical model of hadronization. By using a suitable transition 
operator for hadronization we are able to recover the results of the 
statistical theory, particularly the expressions of the rates of different 
channels. Explicit formulae are obtained for the phase space volume or density 
of states of the ideal relativistic gas in quantum statistics which, for large 
volumes, turn to a cluster decomposition whose terms beyond the leading one 
account for Bose-Einstein and Fermi-Dirac correlations. The problem of the 
computation of the microcanonical ensemble and its comparison with the canonical 
one, which will be the main subject of a forthcoming paper, is addressed.  
\PACS{{12.40.Ee}{} \and {05.30.-d}{}} 
} 
\maketitle
%

\section{Introduction}
\label{intro}

The revived interest in the statistical model of hadron production is mainly 
owing to its application to heavy ion collision where an equilibrated source 
of hadrons is expected. This model has given strikingly good results in 
elementary collisions as well \cite{beca} and this finding has triggered some 
debate about their interpretation \cite{diba}. A proposed one is that 
hadronization occurs at some critical energy density \cite{crit} (or maybe 
another related parameter) of a number of massive pre-hadronic colourless 
extended objects (henceforth referred to as {\em clusters}) which are formed 
as a result of the underlying non-perturbative strong-interaction dynamics and 
which thereafter decay coherently into multihadronic states \cite{stock}. In 
this scheme, the single cluster's decay rate into any channel would be determined 
only by its phase space with no special dynamical weight ({\em phase space dominance}). 
Thereby, the observed statistical equilibrium would not be the effect of a 
collisional thermalization process between formed hadrons over long-lived 
extended regions in the final state, rather of equal quantum transition
probabilities from a cluster to all accessible final states. By accessible 
it is meant that one must comprise only those states fulfilling conservation 
laws, i.e. having the same quantum numbers as the initial cluster's. The set
of states with fixed energy-momentum, angular momenta, parity and internal 
charges is defined as {\em microcanonical ensemble}, though the same name 
is usually employed to denote the set of states with fixed energy-momentum 
and internal charges, relaxing angular momentum and parity conservation. 
We will not make any distinction either; whether the constraints of angular 
momenta and parity are meant to be included, it will be clear from the context. 

Although the microcanonical ensemble is the correct statistical ensemble to
use in hadronizing a single cluster, so far all actual data analyses within 
the statistical model have been carried out in the framework of the 
canonical or grand-canonical ensemble of the hadron gas, i.e. with hadronizing 
sources described in terms of a temperature and taking into account the 
conservation of energy-momentum and angular momentum only on average. The 
microcanonical ensemble has been used very seldom \cite{weai,liu}, mainly 
owing to the hard and long computations involved. Indeed, in high energy 
collisions, where many clusters are produced, the use of the canonical ensemble is 
favoured by fluctuations of masses and volumes, which tend to reduce the importance 
of exact conservation of energy and momentum. It is even possible that fluctuations 
make the system of many clusters equivalent (as far as Lorentz invariant quantities, 
such as average multiplicities, are concerned) to a large global cluster obtained 
by ideally clumping them \cite{becapt}. While the canonical ensemble is 
in fact a better and better approximation of the microcanonical one for large values 
of cluster's mass and volume, we have no quantitative estimate of how large they 
ought to be \footnote{Recently, a calculation has been done for pp collisions with 
a restricted set of hadrons \cite{liu}}, so the use of the canonical ensemble is, 
until now, justified by the agreement with the data. On the other hand, it would be 
desirable to have a more precise and quantitative assessment of the goodness 
of the canonical approximation. An explicit calculation of the microcanonical ensemble 
of the hadron gas is also necessary if we want to test the statistical model at 
lower energies (say $\sqrt s < 10$ GeV) where conservation laws are expected to 
play a major role and canonical approximation is not a good one. Furthermore, it 
would be very useful having at our disposal a Monte-Carlo algorithm for 
microcanonical hadronization of single clusters in high energy collisions to be 
used for numerical calculations of quantities for which an analytical expression 
cannot be obtained. Thus, providing a reliable and fast numerical algorithm for the 
calculation of microcanonical ensemble of the hadron gas and comparing the 
results with the canonical approximations are the main goals of this work 
which will be described in two papers.

In this first paper, we will confine ourselves to the analytical development of 
the microcanonical formalism, while the numerical calculations will be the main
subject of the second paper. In fact, another major motivation of this work is
to provide a consistent formulation of the statistical hadronization model starting
from quantum transition probabilities, which is still lacking despite the apparent 
simplicity of the picture and the fact that its foundations were laid more than 
50 years ago \cite{fermi}. The attempts to derive the statistical theory results 
from $S$-matrix under suitable hypotheses, were mainly carried out by Hagedorn 
\cite{hagesmat} and Cerulus \cite{ceru} on the basis of time-reversal 
arguments. However, the reasoning is quite involved in this approach and one
needs separate treatment of dynamical matrix element averaging for multiplicities
and spectra. The inclusion of quantum statistical effects in the microcanonical 
ensemble of the relativistic hadron gas has been done \cite{hageps} consistently 
only for large volumes and not sufficiently general. 
In fact, in this traditional treatment, particle states confined in the cluster's 
volume are assumed to be eigenstates of energy-momentum, which is true only if the 
volume is so large that the entailed energy-momentum uncertainty can be neglected, 
what is not generally the case when dealing with small volumes (this will
be discussed more in detail at the end of Sect. 2).
Furthermore, in that approach, the whole treatment did not start from the 
statistical theory of multiple production and it is thus not 
easy to generalize if angular momentum and parity conservation are to be 
included.  

Therefore, we believe that a coherent general reformulation of the statistical 
model of hadronization is needed. In this paper, we will go along the whole 
formalism starting from the basic assumptions and will recover some well known
formulae in literature, like those in ref.~\cite{hageps}, as approximations of 
more general ones in case of sufficiently large volumes. In particular, we
will recover the $N$-body relativistic phase space expression (without angular
momentum and parity conservation) without treating confined particle states 
as energy-momentum eigenstates, an assumption which is correct only asymptotically.
We will show how conservation laws are to be implemented in the most general
case, thus providing an usable framework to obtain more general expression of 
$N$-body phase space when angular momentum and other conserved quantities are 
to be taken into account. Furthermore, 
we will explicitely show how the microcanonical ensemble reduces to the canonical 
one for large cluster's volume and mass. The formulae presented in this work 
will be then the basis of the numerical computations in the second paper \cite{micro2}.    

The paper is organized as follows: in Sect.~2 we will introduce a basic formulation 
of the statistical hadronization model and follow the path leading to the microcanonical 
ensemble; in Sect.~3 we will develop in detail the microcanonical formalism for an 
ideal hadron-resonance gas with full quantum statistics; in Sect.~4 the microcanonical 
partition function will be calculated and the approximations needed to obtain closed 
expressions stressed, while in Sect.~5 the transition from the microcanonical to the 
canonical ensemble described; finally, in Sect.~6 the calculation of the physical 
observables will be discussed.

\section{Statistical hadronization of a cluster}

The fundamental assumption of the statistical hadronization model is that the 
final stage of a high energy collision results in the formation of a set of 
extended colourless massive objects, the {\em clusters} or {\em fireballs}, 
producing hadrons in a purely statistical manner: that is, all multihadronic 
states within the cluster volume and compatible with cluster's quantum numbers
are equally likely. Clusters can indeed be thought as very short-lived extended 
resonances, much alike to bags of the bag model \cite{bag}. They differ from
clusters proposed in other hadronization models \cite{herwig} as they are endowed 
with a spacial extension. In this picture,
the cluster's decay rate into a given $N$-particle channel should be proportional 
to the number of multiparticle states within the volume $V$ of the cluster, which 
can be expressed, in the limit of Boltzmann statistics and neglecting angular 
momentum and parity conservation, as:  
\begin{equation}\label{ips}
\Gamma \propto V^N \int \d^3 {\rm p}_1 \ldots \d^3 {\rm p}_N \; \delta^4(P - 
\sum_{i=1}^N p_i)
\end{equation}
where $P$ is the four-momentum of the cluster. 
The distinctive feature of the statistical model is essentially the appearance
of a finite volume in the decay rate, which makes the above expression different 
from that of the decay rate of a massive particle usually found in textbooks,
where asymptotic states are defined over an infinitely large volume:
\begin{equation}\label{ims}
\Gamma \propto \int \frac{\d^3 {\rm p}_1}{2 \epsilon_1} \ldots 
\frac{\d^3 {\rm p}_N}{2 \epsilon_N} \; | M_{if}|^2 \delta^4(P - 
\sum_{i=1}^N p_i)
\end{equation}
While these two equations have in principle a different physical meaning, there
have been several attempts to derive an equation like (\ref{ips}) from 
Eq.~(\ref{ims}) (see e.g. ref.~\cite{hagesmat}). Instead of establishing a 
link between them through a suitable choice of the squared matrix element 
$|M_{if}|^2$, we will try to obtain the formula~(\ref{ips}) starting from a 
suitable {\em ansatz} which will enable us also to recover quantum statistics 
effects (Bose-Einstein or Fermi-Dirac correlations) in a natural way. This 
can be accomplished because a finite volume is involved in Eq.~(\ref{ips})
unlike in Eq.~(\ref{ims}). 

We assume that, as a result of non perturbative QCD-driven evolution, the cluster 
state develops uniform projections over the multihadronic Fock space states 
defined by its volume and compatible with its quantum numbers. Thus, if 
$\vert i \rangle$ is a properly normalized asymptotic state characterized by 
the mass, spin and quantum numbers of the cluster and $\langle f \vert$ 
an asymptotic multihadronic final state, the rate $\Gamma_f$ into the final state 
$f$ is written as:
\begin{equation}\label{prob}
 \Gamma_f = \vert \langle f | W | i \rangle \vert^2 
\end{equation}
where $W$ is an effective transition operator proportional to the projector
over the Hilbert subspace defined by all stationary multihadronic states 
$\vert h_V \rangle$ {\em within the cluster}, namely:
\begin{equation}\label{trans}
 W = \sum_{h_V} \vert h_V \rangle \langle h_V \vert \hat\eta \equiv P_V \hat \eta
\end{equation}
where $\hat\eta$ is an operator depending on strong interaction symmetry group 
invariants (Casimir operators) such as mass, spin, isospin, charge etc. The
state $| h_V \rangle$ will be assumed as a confined stationary free particle 
state within the cluster, with fixed or periodic boundary conditions; the inclusion 
of all resonances as independent states allows to take into account a part of the 
interaction between strongly stable hadrons and this is the reason of the usual 
expression {\em ideal hadron-resonance gas} \cite{hage}. 
 
The operator $W$ is a peculiar one because it is dependent on the shape and 
volume of the cluster, which in fact pertain to the initial conditions. If 
cluster's quantum numbers coincide with those of the initial colliding system, 
(only one cluster is produced) $W$ should commute with all conserved quantities 
in strong interaction to ensure the due selection rules, though this may not 
be necessary if many clusters are produced. 

The commutation requirement is fulfilled for all internal symmetries, charge 
conjugation and for angular momentum and parity provided that the cluster has 
spherical shape (see Appendix A). On the other hand, $W$ does not commute with
energy and momentum as translational symmetry is broken by the assumption
of a finite volume, hence a violation of energy-momentum conservation of
the order of the inverse of the cluster's linear size is implied. However, 
as it will become clear in the following, momentum-integrated rates in 
fact get contribution only from states fulfilling energy-momentum conservation; 
otherwise stated, finite volume introduces a smearing effect on energy and 
momentum which is washed out after kinematical integrations. 
It should also be pointed out that viewing a short-lived object such as a
cluster as an asymptotic state with definite total energy and momentum is 
certainly an approximation and a slight violation of energy-momentum conservation 
is not to be taken as a serious awkwardness. Problems may arise only in 
handling single-cluster collision events, where final states must have the 
energy and momentum of the colliding system.

The Eq.~(\ref{prob}) can be written as:
\begin{eqnarray}\label{prob2}
 \Gamma_f &=& \langle f | W | i \rangle  \langle i | W^\dag | f \rangle =
 \langle f | W {\sf P}_i W^\dag | f \rangle \nonumber \\
 &=& |\eta_i|^2 \langle f | P_V {\sf P}_i P_V^\dag | f \rangle
\end{eqnarray}
where ${\sf P}_i$ is the projector over the initial quantum state and $\eta_i$ is
such that $\hat\eta | i \rangle = \eta_i | i \rangle$. In principle, the 
projection is to be carried out onto a state with definite energy, momentum, 
spin (the Pauli-Lubanski vector), parity, C-parity (if the cluster is neutral) and 
internal charges. Hence the most general projector to be considered reads:
\begin{equation}
   {\sf P}_i = {\sf P}_{P,J,\lambda,\pi} {\sf P}_\chi {\sf P}_{I,I_3} {\sf P}_\Qz  
\end{equation}
where $P$ is the four-momentum of the cluster, $J$ the spin, $\lambda$ the 
helicity, $\pi$ the parity, $\chi$ the C-parity, $I$ and $I_3$ the isospin and 
its third component and $\Qz = (Q_1,\ldots,Q_M)$ a set of $M$ abelian (i.e.
additive) charges such as baryon number, strangeness, electric charge etc.
Of course, the projection ${\sf P}_\chi$ makes sense only if $I_3 = 0$ and
$\Qz = {\bf 0}$; in this case, ${\sf P}_\chi$ commutes with all other projectors. 
 
A state with definite four-momentum, spin, helicity and parity tranforms according 
to an irreducible unitary representation $\nu$ of the extendend Poincar\'e group 
IO(1,3)$^\uparrow$, and the projector ${\sf P}_{P,J,\lambda,\pi}$ can be written by 
using the invariant, suitably normalized, measure $\mu$ as:
\begin{equation}\label{proj}
 {\sf P}_{P,J,\lambda,\pi} = \frac{1}{2} \sum_{z={\sf I,\Pi}} \dim \nu 
 \int \d \mu(g_z) \; D^{\nu \dag}(g_z)^i_i \, U(g_z)   
\end{equation} 
where $z$ is the identity or space inversion ${\sf \Pi}$, $g_z \in 
{\rm IO}(1,3)^\uparrow_{\pm}$, $D^\nu(g_z)$ is the matrix of the irreducible 
representation $\nu$ the initial state $i$ belongs to, and $U(g_z)$ is the unitary 
representation of $g_z$ in the Hilbert space. Similar integral expressions can be 
written for the projectors onto internal charges, for the groups SU(2) (isospin) 
and U(1) (for additive charges). Although projection operators cannot be rigorously
defined for non-compact groups, such as Poincar\'e group, we will maintain this naming 
relaxing mathematical rigour. In fact, for non compact-groups, the projection 
operators cannot be properly normalized so as to ${\sf P}^2 = {\sf P}$ and this 
is indeed related to the fact that $| i \rangle $ has infinite norm. Still, 
we will not be concerned with such drawbacks thereafter, whilst it will be favourable 
to keep the projector formalism. Working in the rest frame of the cluster, with 
$P = (M, {\bf 0})$, the matrix element $D^{\nu\dag}(g_z)^i_i$ vanishes unless the 
Lorentz transformations are pure rotations and this implies the reduction 
of the integration in (\ref{proj}) from IO(1,3)$^\uparrow$ to the subgroup 
${\rm T}(4) \otimes {\rm SU}(2) \otimes {\rm Z}_2$ (see Appendix B). Altogether, 
the projector ${\sf P}_{P,J,\lambda,\pi}$ reduces to:
\begin{eqnarray}\label{proj2}
\!\!\!\!\!\! {\sf P}_{P,J,\lambda,\pi} =&& \frac{1}{(2\pi)^4} \int \d^4 x \; 
\e^{\i P \cdot x} U({\sf T}(x)) \nonumber \\
\!\!\!\!\!\! && \times (2J + 1) \int \d{\sf R} \; D^J({\sf R})^{\lambda *}_\lambda 
 \, U({\sf R}) \, \frac{{\sf I} + \pi U({\sf \Pi})}{2}  
\end{eqnarray} 
$\d {\sf R}$ being the invariant SU(2) measure normalized to 1. The invariant
measure $\d^4 x$ of the translation subgroup has been normalized with a coefficient
$1/(2\pi)^4$ in order to lead to a Dirac delta, as shown below. This is indeed
the general expression of the projector defining the proper microcanonical 
ensemble, where all conservation laws related to space-time symmetries are 
fulfilled.

Hereafter, we will confine ourselves to clusters with fixed energy, momentum and
abelian charges while conservation of angular momentum, isospin, parity and C-parity 
will be disregarded. This is expected to be a very good approximation in high energy 
collisions, where many clusters are formed and these latter constraints should not 
play a significant role \cite{hage,ceru}. On the other hand, they cannot be disregarded 
in very small hadronizing systems (e.g. \ppb at rest \cite{heinz}) and, in such
circumstances, the full projection operation in Eq.~(\ref{proj2}) should be carried 
out. 
As has been mentioned in the introduction, the set of states with fixed values of 
energy, momentum and abelian charges is defined microcanonical ensemble as well and 
we will stick to this convention. 
 
Dealing with clusters with an unspecified value of angular momentum isospin and parity
means, from a statistical mechanics point of view, that all possible projections over 
definite values of those quantum numbers occur with their statistical weight. In other 
words, we shall sum over all $J,\lambda,I,I_3,\pi,\chi$, which amounts to simply remove 
the relevant projection operators in virtue of the completeness relations such as, for 
instance:
\begin{equation}
 \sum_{J \lambda} (2J + 1) \int \d{\sf R} \; 
 D^J({\sf R})^{\lambda *}_\lambda \, U({\sf R}) = {\sf I}    
\end{equation}
In this case, the projector operator onto the initial state reduces to the more 
familiar form:
\begin{eqnarray}\label{proj3}
&& \!\!\!\!\!\!\! \!\! {\sf P}_i \rightarrow {\sf P}_P {\sf P}_\Qz =  
\frac{1}{(2\pi)^4} \int \d^4 x \; \e^{\i P \cdot x} \e^{-\i P_{\rm op} \cdot x} 
\nonumber \\
&&  \!\!\!\!\!\!\!\!\! \times \frac{1}{(2\pi)^M} \int_{-\pivs}^{+\pivs} \!\!\!\! 
\d^M \phi \; \e^{\i \Qz \cdot \phivs} \e^{-\i \Qz_{\rm op} \cdot \phivs} = 
\delta^4 (P - P_{\rm op}) \, \delta_{\Qz,\Qz_{\rm op}}
\end{eqnarray}
where $\pivs = (\pi,\ldots,\pi)$ and the group generators $P_{\rm op}$ and 
$\Qz_{\rm op}$ have been introduced. The appearance of Dirac and Kronecker deltas 
in Eq.~(\ref{proj3}) reflects the abelian nature of the leftover space-time 
translations and U(1) groups. 
By using the latter expression of the projector ${\sf P}_i$, the Eq.~(\ref{trans}) 
and inserting two identity resolutions, Eq.~(\ref{prob2}) turns to :
\begin{eqnarray}\label{projt}
\Gamma_f =&& |\eta_i|^2 \sum_{h_V h'_V} \sum_{f' f''} \langle f | h_V \rangle 
\langle h_V | f' \rangle \langle f'' | h'_V \rangle \langle h'_V | f \rangle \nonumber \\
&& \times \langle f' | \delta^4 (P - P_{\rm op}) \, \delta_{\Qz,\Qz_{\rm op}} | f'' \rangle 
\end{eqnarray}
and, taking $f', f''$ states as energy-momentum and charges eigenstates:
\begin{eqnarray}\label{rate}
\Gamma_f = && \sum_{h_V h'_V} \sum_{f'} \langle f | h_V \rangle 
\langle h_V | f' \rangle \langle f' | h'_V \rangle \langle h'_V | f \rangle \nonumber \\
&& \times \delta^4 (P - P_{f'}) \, \delta_{\Qz,\Qz_{f'}} 
\end{eqnarray}
that is:
\begin{equation}\label{ratebis}
\Gamma_f = |\eta_i|^2 \sum_{f'} \Big| \sum_{h_V} \langle f | h_V \rangle 
\langle h_V | f' \rangle \Big|^2  \delta^4 (P - P_{f'}) \, \delta_{\Qz,\Qz_{f'}}
\end{equation}
Multiparticle states in the Fock space are characterized by a set of integer 
occupation numbers for all the species and for all the kinematical states. 
This also applies to the general state $| h_V \rangle$ as long as it represents,
as it has been assumed, free hadron and resonance states within the cluster, 
so one can write $| h_V \rangle = |\{\tilde N_j \} k_V \rangle$ where $\{\tilde N_j \} = 
(\tilde N_1,\ldots,\tilde N_K)$ is a $K$-uple of integer numbers one for each
hadron species $j$ and $k_V$  
denotes a set of kinematical variables, depending on the spacial region with
volume $V$, describing the state of the $\tilde N=\tilde N_1+\tilde N_2+\ldots+\tilde N_K$ 
particles. Similarly, we can rewrite the states belonging to the complete basis 
as $| f \rangle = |\{ N_j \} k \rangle$ where now $k$ is meant to be a set 
of proper momenta and polarizations. Note that the expression~(\ref{ratebis}) allows
transitions to states $| f \rangle$ with energy-momentum different from $P$, 
unless the volume is infinitely large. This tells us, as has been mentioned, that 
the energy-momentum spread is of the order of the inverse of the cluster's 
linear size.   

To further develop Eq.~(\ref{ratebis}) we shall assume that:
\begin{equation}\label{orth}
\langle \{ \tilde N_j \} k_V | \{ N_j \} k \rangle = 0 
\qquad {\rm if} \; \tilde N_j \ne N_j \qquad \forall j
\end{equation}
Hence, it is required that states with different particle composition, either 
within the bounded region or in the whole space, are orthogonal. Indeed, there are 
two contraindications to this assumption. The first is of a more fundamental
character: in relativistic quantum field theory a condition like (\ref{orth}) cannot 
be exactly true as stationary states localized in a finite region are not eigenstates 
of the properly defined particle number operator (localization involves the creation 
of particle-antiparticle pairs). However, this effect is relevant if the size of the 
region is lower than the Compton wavelenght of the particle $1/m$, which is at most 
(for pions) $\approx 1.4$ fm, corresponding to a volume of $\approx 3$ fm$^3$; for all 
other hadrons, this volume is significantly smaller. Henceforth, we will assume that 
volumes to be dealt with are larger (not too much though) and will take a 
non-relativistic quantum mechanical treatment as a good approximation. The second
is concerned with strongly decaying resonances, which, in principle, should not be 
orthogonal to the states of their decay products; however, we have assumed that
resonances are to be treated as independent states, so the orthogonality relation 
is correct in the framework of the ideal hadron-resonance gas.   
       
With these two caveats in mind, we proceed to calculate the total rate of some 
{\em channel}, i.e. a multihadronic configuration $\{ N_j\}$, by summing over the physical 
observables $k$, being $|f\rangle = |\{ N_j\} k \rangle$. Applying the sum
to the right hand side of Eq.~(\ref{rate}), taking into account the condition
(\ref{orth}) and the completeness of the set $|\{ N_j\} k \rangle$, one obtains: 
\begin{eqnarray}
&& \!\!\! \sum_{k} \langle f | h_V \rangle \langle h'_V | f \rangle = 
\prod_j \delta_{N_j\tilde N_j} \delta_{N_j \tilde N'_j}
\sum_k \langle \{ N_j \} k | \{ N_j \} k_V \rangle \nonumber \\
&& \!\!\! \times \langle \{ N_j \} k'_V | \{ N_j \} k \rangle 
= \langle h_V | h'_V \rangle \prod_j \delta_{N_j\tilde N_j} \!\! 
= \delta_{h_V h'_V} \prod_j \delta_{N_j\tilde N_j} \nonumber \\
&&
\end{eqnarray}
and, therefore:
\begin{eqnarray}\label{rate2}
&& \!\!\!\!\!\!\!\!\!\! \Gamma_{\Nj} \equiv \sum_k \Gamma_{\Nj,k} \nonumber \\
&& \!\!\!\!\!\!\!\!\!\! = |\eta_i|^2 \sum_{k'} \sum_{k_V} | \langle \{ N_j \} k' | 
\{ N_j \} k_V \rangle |^2  \delta^4 (P - P_{f'}) \, \delta_{\Qz,\Qz_{f'}}
\end{eqnarray}
The above equation~(and, maybe more apparently, Eq.~(\ref{basic}) below) shows 
that only kinematical states fulfilling energy-momentum conservation contribute 
to the total rate of a channel even though the transition to final states with 
$P_f \neq P$ is allowed, as discussed. 

We are now in a position to recover an expression like~(\ref{ips}) mentioned at
the beginning of this section. More specifically, we can prove that the right 
hand side in Eq.~(\ref{rate2}) is $|\eta_i|^2$ times 
the usual expression of the probability of the multihadronic configuration $\Nj$ to 
occur in the microcanonical ensemble of an ideal hadron-resonance gas with 
four-momentum $P$, charges $\Qz$ and volume $V$, as long as the aforementiond
relativistic quantum field effects are disregarded. We will show this first in the 
simple case of a channel with all different particles, i.e. $N_j \le 1 \; \forall j$; 
the case of identical particles will be handled in the next section. The scalar 
product in Eq.~(\ref{rate2}) factorizes, so that: 
\begin{equation}\label{factor}
\sum_{k_V} | \langle \{ N_j \} k' | \{ N_j \} k_V \rangle |^2 =
\prod_{i=1}^N \sum_{{\bf k}_i,\tau_i} | \langle {\bf p}'_i \sigma'_i | 
{\bf k}_i \tau_i \rangle |^2 
\end{equation}
where $i=1,\ldots,N$ is the single-particle index. The variable ${\bf p}$ is
a momentum whilst ${\bf k}$ denotes three variables defining the state
of the particle within the region with volume $V$ (e.g. a plane wave vector for a
rectangular box or energy and angular momenta for a sphere). The variables 
$\sigma'_i$ and $\tau_i$ labels different polarization states of the particle 
and may refer to different projections of the spin (or the helicity); we will 
assume that the transformation from $\tau$ to $\sigma$ is unitary. 
As long as $|{\bf k}, \tau\rangle$ is a complete set of one-particle states
in the region with volume $V$, as a consequence of the completeness of the 
states $| h_V \rangle$, it can be shown that, in the non-relativistic quantum
mechanics approximation (see Appendix C):
\begin{equation}\label{volume}
\sum_{{\bf k},\tau} | \langle {\bf p} \sigma | {\bf k} \tau \rangle |^2 = 
\frac{V}{(2\pi)^3} 
\end{equation}
Thus, taking into account that $\sum_{k'}= [\prod_{i=1}^N \sum_{\sigma_i} 
\int \d^3 {\rm p}_i]$ \footnote{The notation 
$[\prod_i \int \d^3 {\rm p}_i]$ stands for the integral operator $\int \d^3 
{\rm p}_1 \ldots \int \d^3 {\rm p}_N$ and it is understood to act on its right hand 
side argument} and Eq.~(\ref{factor}), Eq.~(\ref{rate2}) becomes: 
\begin{eqnarray}\label{basic}
\Gamma_{\Nj} &=& |\eta_i|^2 \frac{V^N}{(2 \pi)^{3N}} \Big[ \prod_{i=1}^N (2J_i+1) 
\int \d^3 {\rm p}_i \Big] \; \delta^4 (P-\sum_i p_i) \nonumber \\
&=& |\eta_i|^2 \Omega_{\Nj} 
\end{eqnarray}
where charge conservation $\sum_j N_j \qj = \Qz$ is understood. 
Therefore, the rate $\Gamma_{\Nj}$ is proportional to the usual expression of 
the {\em phase space volume} or density of states per four-momentum cell 
$\Omega_{\Nj}$ of the multihadronic configuration $\Nj$ in the microcanonical 
ensemble of the ideal hadron-resonance gas. It is to be emphasized that this 
formula, which has been used by many authors in the framework of the statistical 
model, is not the most general though, as all particles must be different.
Therefore, it corresponds to assuming the classical Boltzmann statistics.  
We will see in the next section that, if quantum statistics are taken into 
account, the integral in Eq.~(\ref{basic}) is indeed a single term of an 
expansion. 

Even though our derivation might look unnecessarily elaborated, the $N$-body
relativistic phase space volume in Boltzmann statistics Eq.~(\ref{basic})
has been recovered starting from purposely built quantum mechanical transition 
probabilities without invoking any time reversal argument or averaging procedures 
like in previous treatments \cite{hagesmat,ceru}. Furthermore, it should be
emphasized that this derivation is more general than previous treatments because 
we did not consider particle states within the cluster as energy-momentum 
eigenstates. In fact, Eq.~(\ref{basic}) is obtained in a traditional approach 
\cite{hageps} by working out the expression:
\begin{equation}\label{hpspace}
\Omega_{\Nj} = \sum_{\rm states} \delta^4(P - P_{\rm state}) 
 \delta_{\Nj,\Nj_{\rm state}}
\end{equation}
with the key assumption that particles within the cluster have indeed definite 
four-momenta and so does the whole multiparticle state:
\begin{equation}\label{pstate}
 P_{\rm state} = \sum_i p_{i} \qquad ;
\end{equation}
and finally replacing the sum over particle states in the cluster with an integration:
\begin{equation}\label{cell}
 \sum_{\bf k} \rightarrow \frac{V}{(2\pi)^3} \int \d^3 {\rm p}
\end{equation}
Altogether, this approach can be a good approximation only for very large volumes, 
because only for large volumes can the uncertainty on energy-momentum entailed by 
localization within the cluster be negligible. For smaller volumes, the localized 
multiparticle states, that we have denoted as $|h_V\rangle$, are not eigenstates 
of energy-momentum and their spread in energy-momentum cannot be neglected. 
It is worth making a rough
estimate of how large the volume ought to be for the traditional approach to be 
valid. This can done in two ways: requiring that the uncertainty in momentum 
for a single-particle localized state is not larger than order of, say, 10\%
or arguing that the approximation~(\ref{cell}) is indeed a good one 
provided that the number of phase space cells is at least of the order of 10-100. 
Working out (\ref{cell}) or using the indeterminacy principle, it turns out in
both cases that the cluster's linear size should be larger than $\approx (6$-$10)/p$, 
where $p$ is the typical momentum of the particles at hadronization. Since this 
is of the order of some hundreds MeV, the linear size must be of the order of, 
say, 3-10 fm, which is consistently larger than the limit set by the aforementioned 
condition on the Compton wavelength of the particles, of the order of a fraction 
of fermi. Therefore, the requirement on the volume for the validity of the traditional
treatment is more stringent than that needed for the present one.

Although in the case of Boltzmann statistics, the traditional and the present
approach lead to the same expression Eq.~(\ref{basic}) for the $N$-body relativistic
phase space volume, different expressions are found in the case of quantum 
statistics, that is with identical particles, as it will be shown in the next 
section.

\section{Identical particles and cluster decomposition}

If there are identical particles in the channel, the equation~(\ref{rate2}) 
holds but Eq.~(\ref{factor}) does not and has to be modified. For sake of 
simplicity, we will start with the case of only one kind of particle in the channel 
and assume that charge conservation is fulfilled. As has already mentioned, 
relativistic quantum field effects will be disregarded, namley cluster's size is 
assumed to be significantly larger than the Compton wavelenght of the particle. 
The correspondance between Fock space and multiparticle tensor space requires the 
identification:
\begin{eqnarray}\label{tensor}
&& \!\!\!\!\!\!\!\!\!\! | \Nj k_V \rangle \rightarrow \nonumber \\
&& \!\!\!\!\!\!\!\!\!\! \sum_{p}\chi(p)^b \frac{1}{\sqrt{N!n_1!\ldots n_M!}}
|{\bf k}_{p(1)}\tau_{p(1)},\ldots,{\bf k}_{p(N)} \tau_{p(N)} \rangle 
\end{eqnarray}
where $p$ is a permutation of the integers $1,\ldots,N$ and $\chi(p)$ its parity; 
the $n_i$'s are the number of times a given vector ${\bf k}_i$ recurs in the state 
with $\sum_{i=1}^M n_i = N$; $b=0$ for bosons and $b=1$ for fermions. As there is 
only one particle species, the phase space volume $\Omega_{\Nj}$ can be denoted 
with $\Omega_N$ and can be calculated by using Eq.~(\ref{rate2}). 
Replacing $| \Nj k'\rangle$ with:
\begin{equation}\label{tensor2}
 | \Nj k \rangle \rightarrow  \sum_{p} \frac{\chi(p)^b }{\sqrt N!}
|{\bf p}_{p(1)}\sigma_{p(1)},\ldots,{\bf p}_{p(N)} \sigma_{p(N)} \rangle 
\end{equation}
similarly to Eq.~(\ref{tensor}), and dividing by $1/N!$ in order to avoid multiple
counting of (anti-)symmetric basis tensors when integrating over all possible 
momenta, we find:
\begin{eqnarray}\label{rateid}
&& \!\!\!\! \Omega_N = \frac{\Gamma_N}{|\eta_i|^2} = \Big[ \prod_{i=1}^N \sum_{\sigma_i} 
\int \d^3 {\rm p}_i \Big] \delta^4 (P-P_f) \nonumber \\
&& \!\!\!\! \times \sum_{k_V} \Big| \sum_{p} \chi(p)^b \frac{1}{\sqrt{N!n_1!\ldots n_M!}}
\langle {\bf p}_1\sigma_1,\ldots | {\bf k}_{p(1)} \tau_{p(1)},\ldots \rangle \Big|^2 
\nonumber \\
&&
\end{eqnarray}
In Eq.~(\ref{rateid}) and hereafter $P_f$ must be understood as the sum of the 
four-momenta of all particles in the channel. The last factor in the above equation 
can be worked out as follows:
\begin{eqnarray}
&& \sum_{k_V} \Big| \sum_{p} \chi(p)^b \frac{1}{\sqrt{N!n_1!\ldots n_M!}}
\langle {\bf p}_1\sigma_1,\ldots | {\bf k}_{p(1)} \tau_{p(1)},\ldots \rangle \Big|^2 
\nonumber \\
&& = \sum_{k_V} \frac{1}{N! n_1!\ldots n_M!} \sum_{p,q} \chi(p)^b\chi(q)^b \nonumber \\
&& \times \langle {\bf p}_1\sigma_1,\ldots | {\bf k}_{p(1)} \tau_{p(1)},\ldots \rangle 
\langle {\bf k}_{q(1)}\tau_{q(1)},\ldots | {\bf p}_1 \sigma_1,\ldots \rangle
\nonumber \\
&& = \frac{1}{N!^2} \sum_{p,q} \chi(pq)^b \prod_{i=1}^N \sum_{{\bf k}_i \tau_i} 
\langle {\bf p}_{p(i)} \sigma_{p(i)} | {\bf k}_i \tau_i \rangle 
\langle {\bf k}_i \tau_i | {\bf p}_{q(i)} \sigma_{q(i)}\rangle \nonumber \\
&&
\end{eqnarray}
where, in the last equality, we have redefined the dummy permutation indices $p,q$ 
as their inverse and multiplied each term by a factor $n_1!\ldots n_M!/N!$ in order 
to avoid multiple counting of the symmetric (antisymmetric) basis tensors 
$| h_V \rangle$ when the sum over all possible vectors ${\bf k}$ and polarizations 
$\tau$ is carried out. Finally, taking into account that also ${\bf k}_i$ and $\tau_i$ 
are dummy indices, one sum over permutations can be trivially performed and we are 
left with the transformation: 
\begin{eqnarray}\label{sumperm}
&& \sum_k \sum_{k_V} | \langle \{ N_j \} k | \{ N_j \} k_V \rangle |^2 \rightarrow 
\frac{1}{N!} \sum_{r} \chi(r)^b \nonumber \\
&& \times \Big[ \prod_{i=1}^N \int \d^3 {\rm p}_i \Big]
\sum_{{\bf k}_i \tau_i} \langle {\bf p}_i \sigma_i | {\bf k}_i \tau_i \rangle 
\langle {\bf k}_i \tau_i | {\bf p}_{r(i)} \sigma_{r(i)}\rangle
\end{eqnarray}
being $r=p^{-1}q$ and $\chi(r)=\chi(p^{-1}q)=\chi(p^{-1})\chi(q)=\chi(p)\chi(q)$.
The inner sums in the above equality yield (see Appendix C):
\begin{equation}\label{bec}
\sum_{{\bf k}_i \tau_i} \langle {\bf p}_i \sigma_i | {\bf k}_i \tau_i \rangle 
\langle {\bf k}_i \tau_i | {\bf p}_{r(i)} \sigma_{r(i)}\rangle 
= \frac{\delta^{\sigma_i}_{\sigma_{r(i)}}}{(2\pi)^3} \int_V \!\! \d^3 {\rm x} \; 
\e^{\i {\bf x \cdot}({\bf p}_{r(i)}-{\bf p}_i)}
\end{equation}
so the following expression of the phase space volume $\Omega_N$ for $N$ identical 
particles is obtained:
\begin{eqnarray}\label{rateid2}
\Omega_N =&& \sum_r \frac{\chi(r)^b}{N!} \sum_{\sigma_1,\ldots,\sigma_N} 
\int \d^3 {\rm p}_1 \ldots \d^3 {\rm p}_N \; \delta^4 (P-P_f) \nonumber \\
&& \times  \prod_{i=1}^N \delta^{\sigma_i}_{\sigma_{r(i)}} \frac{1}{(2\pi)^3} 
\int_V \d^3 {\rm x} \; \e^{\i {\bf x \cdot}({\bf p}_{r(i)}-{\bf p}_i)} 
\end{eqnarray}
Hence, the phase space volume of $N$ identical particles is given by the sum of $N!$ 
terms and it is thus enhanced or suppressed with respect to the case of 
distinguishable particles. As it will be proved in the following, this effect 
is owing to the finite volume and, thereby, this model naturally accounts for
Bose-Einstein and Fermi-Dirac correlations. 

To develop Eq.~(\ref{rateid2}), it is useful to recall that any permutation $r$ 
of $N$ integers can be uniquely decomposed into the product of cyclic permutations, 
that is $r = c_1 \ldots c_H$. Let $n$ be the number of integers in each cyclic permutation 
and let $h_n$ be the number of cyclic permutations with $n$ elements in $r$ so that 
$\sum_{n=1}^\infty n h_n = N$. The set of integers $h_1,\ldots,h_N \equiv \{h_n\}$,
with $\sum_{n=1}^\infty h_n \equiv H$, is usually defined as a {\em partition} 
and different permutations having the same structure of cyclic decomposition, that 
is the same number of integers for each $c_l$ (i.e. the same partition), belong to 
the same conjugacy class of the permutation group S$_N$. The crucial 
observation is that each term in Eq.~(\ref{rateid2}) is invariant over a conjugacy 
class, or, in other words, depends only on the partition $\{h_n\}$; this happens 
because different permutations in the same conjugacy class differ only by a 
redefinition of the integers $1,\ldots,N$ and this is just a change of the name of 
the dummy integration variables and sum indices in Eq.~(\ref{rateid2}). The number 
of permutations of S$_N$ belonging to a given conjugacy class is a well known number 
\cite{weyl}, namely $N!/\prod_{n=1}^N n^{h_n} h_n!$. Furthermore, for the representative 
cyclic permutation $c$ with $n$ elements $(1,\ldots,n)$:
\begin{equation}
\sum_{\sigma_1,\ldots,\sigma_n} \prod_{i=1}^n \delta^{\sigma_i}_{\sigma_{c(i)}} =
 (2J+1) 
\end{equation}
so that:
\begin{equation}
\sum_{\sigma_1,\ldots,\sigma_N} \prod_{i=1}^N \delta^{\sigma_i}_{\sigma_{r(i)}} =
 \prod_{l=1}^H \sum_{\sigma_1,\ldots,\sigma_{n_l}} \prod_{i_l=1}^{n_l} 
\delta^{\sigma_{i_l}}_{\sigma_{c_l(i_l)}} = (2 J +1 )^H  
\end{equation}
where $r=c_1\ldots c_H$ and $n_l$ is the number of integers in the cyclic permutation
$c_l$. By defining:
\begin{equation}\label{fint}
 F_{n_l} = \prod_{i_l=1}^{n_l} \frac{1}{(2\pi)^3} \int_V \d^3 {\rm x} \; 
 \e^{\i {\bf x \cdot}({\bf p}_{c_l(i_l)}-{\bf p}_{i_l})}
\end{equation}
and taking into account that $\chi(c_l)=(-1)^{n_l+1}$, we can finally rewrite 
Eq.~(\ref{rateid2}) as:
\begin{eqnarray}\label{rateid3}
 && \Omega_N = \sum_{\hpart} \Omega_N (\hpart) = \nonumber \\
 && \sum_{\hpart} \frac{(\mp 1)^{N + H}(2J + 1)^H }
  {\prod_{n=1}^N n^{h_n} h_n!} \Big[\prod_{i=1}^N \int \d^3 {\rm p}_i\Big]
 \delta^4 (P-P_f) \prod_{l=1}^H F_{n_l} \nonumber \\
\end{eqnarray}
where the upper sign applies to fermions, the lower to bosons.
Therefore, the phase space volume for a channel with $N$ identical particles
consists of a large number of terms $\Omega_N (\hpart)$, each corresponding 
to a partition $\hpart$, what is usually called in statistical mechanics 
{\em cluster decomposition}\footnote{The term cluster in this context has
nothing to do with our previous definition of an individual hadronizing source}. 

In the large volume limit, the dominant term is the one with the highest power 
of $V$ and this corresponds to the partition $(h_1,h_2,\ldots,h_N)=(N,0,\ldots,0)$, 
i.e. the identical permutation. In this case there are $N$ factors $F_{1}=V/(2\pi)^3$ 
and the whole term reads:
\begin{eqnarray}\label{leading}
 && \Omega_N(N,0,\ldots,0) = \nonumber \\
 && \Big[\frac{V(2J+1)}{(2\pi)^3}\Big]^N \frac{1}{N!}
 \int \d^3 {\rm p}_1 \ldots \d^3 {\rm p}_N  \; \delta^4 (P-P_f) 
\end{eqnarray} 
which is the phase space volume for a set of $N$ identical particles in the classical
Boltzmann statistics; we have indeed recovered the phase space volume quoted in 
Eq.~(\ref{basic}). All other terms of the expansion in Eq.~(\ref{rateid3}) 
have a lower power of $V$. The next-to-leading term correspond to the conjugacy 
class of permutations with one exchange and $N-2$ unchanged integers, i.e. 
$(h_1,h_2,h_3,\ldots,h_N)=(N-2,1,0,\ldots,0)$
In the large volume limit it is easily seen, looking at Eq.~(\ref{fint}), that 
$F_2 \rightarrow \delta^3({\bf p}_2-{\bf p}_1) V/(2\pi)^3$ and the whole term thus 
reads:
\begin{eqnarray}\label{ntl}
&& \Omega_N (N-2,1,0,\ldots,0) = \frac{(\mp 1)}{2(N-2)!} 
 \Big[\frac{V(2J+1)}{(2\pi)^3}\Big]^{N-1} \nonumber \\
&& \times \int \d^3 {\rm p}_2 \ldots \d^3 {\rm p}_N \; \delta^4 (P-P_f) 
\end{eqnarray} 
where $P_f = 2p_2 + p_3 +\ldots p_N$. Introducing the new integration variables 
${\bf p}'= 2{\bf p}_2$ the energy term $2 \varepsilon_2$ becomes $\sqrt{{\rm p}'^2 +
(2m)^2}$ and Eq.~(\ref{ntl}) can be rewritten as:
\begin{eqnarray}\label{ntl2}
&& \Omega_N (N-2,1,0,\ldots,0) = \frac{(\mp 1)}{2^4(N-2)!} 
 \Big[\frac{V(2J+1)}{(2\pi)^3}\Big]^{N-1} \nonumber \\
&& \times \int \d^3 {\rm p'} \d^3 {\rm p}_3 \ldots \d^3 {\rm p}_N \; 
  \delta^4 (P-p'-\sum_{i=3}^N p_i) 
\end{eqnarray} 
Aside from the sign and an overall normalization factor $1/16$, this term corresponds to
the Boltzmann limit (\ref{ntl}) of the phase space volume for a set of $N-2$ 
identical particles plus a new particle (labelled with a prime) obtained by clumping
particles 1 and 2 into a lump with a mass twice the mass of 1 and 2 and the same spin.

Actually, this kind of interpretation holds for all of the terms in Eq.~(\ref{rateid3}). 
In fact, in the large volume limit, each $F_n$ implies the elimination of $n-1$ integration 
variables through the appearance of Dirac deltas, while a single $V/(2\pi)^3$ factor is
left because of the cyclic structure of the permutation, namely:
\begin{equation}\label{asympt}
 F_{n_l} \rightarrow \frac {V}{(2\pi)^3} \prod_{i_l=1}^{n_l-1}
  \delta^3({\bf p}_{i_l}-{\bf p}_{i_l+1})
\end{equation} 
Then, after trivial integrations are carried out in Eq.~(\ref{rateid3}), the 
Dirac delta forcing conservation of four-momentum turns into $\delta^4 (P_i-
n_1 p_1 - n_2 p_{n_1+1} \ldots - n_H p_{n_{H-1}+1})$ and new integrations variables 
can be introduced:
\begin{equation}\label{lumping}
 {\bf p}'_1 = n_1 {\bf p}_1 \;\; {\bf p}'_2 = n_2 {\bf p}_{n_1+1} \; \ldots \; 
 {\bf p}'_H = n_H {\bf p}_{n_{H-1}+1}      
\end{equation}
as well as new energies:
\begin{equation}
  \varepsilon' = n_l \varepsilon = n_l \sqrt{{\rm p}^2+m^2} = \sqrt{{\rm p}'^2+(n_l m)^2}      
\end{equation}
Therefore, the term corresponding to the partition $\hpart$ can be written as:
\begin{eqnarray}
 \Omega_N(\hpartj) = && 
  \frac{(\mp 1)^{N + H}}{\prod_{n=1}^N n^{h_n} h_n! \prod_{l=1}^H n_l^3} 
 \Big[ \frac{V (2J+1)}{(2\pi)^3} \Big]^H \nonumber \\
&& \times \int \d^3 {\rm p}'_1 \ldots \d^3 {\rm p}'_H \; \delta^4 (P - \sum_{l=1}^H p'_l)  
\end{eqnarray}
where particles are now clumped into $H$ lumps with mass equal to $n_l m$ and spin
$J$. Since $\prod_{l=1}^H n_l^3 = \prod_{n=1}^N n^{3h_n}$, the above equation can
be written also as:
\begin{eqnarray}\label{general}
\Omega_N(\hpartj) = && \frac{(\mp 1)^{N + H}}{\prod_{n=1}^N n^{4h_n} h_n!} 
 \Big[ \frac{V(2J+1)}{(2\pi)^3} \Big]^H \nonumber \\
&& \times \int \d^3 {\rm p}'_1 \ldots \d^3 {\rm p}'_H \; 
 \delta^4 (P - \sum_{l=1}^H p'_{l})  
\end{eqnarray}
We can thus conclude that the general term relevant to the cluster decomposition 
of the phase space volume of a set of $N$ identical particles can be obtained by 
calculating the phase space volume, in the Boltzmann statistics, of a suitable
set of lumps having as mass multiple integer values of $m$ and spin $J$, weighted 
by an overall coefficient of $(\mp 1)^{N-H}/\prod_n n^{4h_n}$. Note that the factors 
$1/h_n!$ already take into account the identity of the lumps. 
 
After having inferred the expressions of the phase space volume of a channel with 
$N$ identical particles, the generalization to a channel $\Nj$ (see Sect.~2) with 
an arbitrary number of groups of identical particles for each species $j$ is rather 
straightforward and can be achieved by going along the previous arguments. Thereby, 
the following equations are obtained which are extensions of Eqs.~(\ref{sumperm}),
(\ref{rateid2}), (\ref{rateid3}) respectively: 
\begin{eqnarray}
&& \!\!\!\!\!\!\!\!\!\!\!\!
\sum_k \sum_{k_V} | \langle \{ N_j \} k | \{ N_j \} k_V \rangle |^2 \rightarrow
\prod_j \frac{1}{N_j!} 
\sum_{r_j \in {\rm S}_{N_j}} \chi(r_j)^{b_j} \nonumber \\
&& \!\!\!\!\!\!\!\!\!\!\!\! 
\times \Big[ \prod_{i_j=1}^{N_j} \d^3 {\rm p}_{i_j} \Big] \!\!
\sum_{{\bf k}_{i_j} \tau_{i_j}} \! \langle {\bf p}_{i_j} \sigma_{i_j} | {\bf k}_{i_j}
\tau_{i_j} \rangle \langle {\bf k}_{i_j} \tau_{i_j} | {\bf p}_{r_j(i_j)} 
\sigma_{r_j(i_j)}\rangle
\end{eqnarray}
,
\begin{eqnarray}
\!\!\!\!\!\!\!\!\!\!\!\! && \Omega_{\Nj}= \sum_{\sigma_1,\ldots,\sigma_N} 
\int \d^3 {\rm p}_1 \ldots \d^3 {\rm p}_N \; \delta^4 (P-P_f) \nonumber \\
\!\!\!\!\!\!\!\!\!\!\!\! && \times \prod_j \sum_{r_j \in {\rm S}_{N_j}} \!\!\! 
  \frac{\chi(r_j)^{b_j}}{N_j!} 
 \prod_{i_j=1}^{N_j} \frac{\delta^{\sigma_{i_j}}_{\sigma_{r_j(i_j)}}}{(2\pi)^3} 
 \int_V \d^3 {\rm x} \; \e^{\i {\bf x \cdot}({\bf p}_{r_j(i_j)}-{\bf p}_{i_j})} 
\end{eqnarray}
and
\begin{eqnarray}\label{clusexp0}
\!\!\!\!\!\!\!\!\Omega_{\Nj} =&& \int \d^3 {\rm p}_1 \ldots \d^3 {\rm p}_N \; 
\delta^4 (P-P_f)  \nonumber \\
\!\!\!\!\!\!\!\! && \times \prod_j \sum_{\hpartj} \frac{(\mp 1)^{N_j + H_j} (2J + 1)^{H_j}}
 {\prod_{n_j=1}^{N_j} n_j^{h_{n_j}} h_{n_j}!} \prod_{l_j=1}^{H_j} F_{n_{l_j}}
\end{eqnarray}
with $H_j = \sum_{n_j=1}^{N_j} h_{n_j}$ and $N_j = \sum_{n_j=1}^{N_j} n_j h_{n_j}$.
The above expression is the most general for the microcanonical phase space volume 
of the multihadronic channel $\Nj$ in the ideal hadron-resonance gas framework
with full quantum statistics and generalizes the expression obtained in 
ref.~\cite{hageps}:
\begin{eqnarray}\label{clusexp}
\!\!\!\!\!\!\!\!\Omega_{\Nj} &=& 
\Bigg[ \prod_j \sum_{\hpartj} (\mp 1)^{N_j + H_j} \frac{1}{\prod_{n_j=1}^{N_j}
 n_j^{4h_{n_j}} h_{n_j}!} \nonumber \\
\!\!\!\!\!\!\!\! && \Big[ \prod_{l_j=1}^{H_j} \frac{V (2J_j+1)}{(2\pi)^3} 
  \int \d^3 {\rm p}'_{l_j}\Big]\Bigg] \delta^4 (P_i - \!\!\! \sum_{j,l_j=1}^{H_j} 
 \!\! p'_{l_j} )  
\end{eqnarray}
where, for a set of partitions ${\{h_{n_1}\}}, \ldots, {\{h_{n_K}\}}$, the 
four-momenta $p'_{l_j}$ are those of lumps of particles of the same species $j$ 
($H_j$ in number) with mass $n_j m_j$ and spin $J_j$.
While Eq.~(\ref{clusexp0}) is indeed the correct expression of the microcanonical 
relativistic phase space volume of the channel $\Nj$ in the statistical hadronization 
model, Eq.~(\ref{clusexp}) turns out to be a special case of Eq.~(\ref{clusexp0})
and in fact can be derived from it by replacing $F_{n_l}$ with their limiting 
expressions~(\ref{asympt}). Thus, the expression~(\ref{clusexp}) is a good 
approximation of~(\ref{clusexp0}) only in the limit of large volumes.  
In fact, the derivation of the relativistic phase space volume in Eq.~(\ref{clusexp})
in ref.~\cite{hageps} was based on the assumption that particle states within the 
cluster are energy-momentum eigenstates, which is a good approximation only for 
large volumes, as extensively discussed at the end of the previous section.
For the expression Eq.~(\ref{clusexp0}) to be valid, we now just need the linear size 
of the cluster must be sufficiently larger than the Compton wavelenght of the 
involved particles in order to neglect relativistic quantum field effects.

The fact that the leading Boltzmann terms in the general expression (\ref{clusexp0}) 
and the approximate one (\ref{clusexp}) are the same, as already pointed out in 
Sect. 2, reduces the actual numerical impact of this generalization on many, yet
not all, observables. For instance, at the actual temperature values of about 160 MeV 
found in previous analyses of many high energy collisions in the canonical ensemble 
\cite{beca,becapt}, quantum statistics corrections on average particle multiplicities 
turned out to be significant for pions only (about 10\%), whilst they can be neglected for 
all other hadrons. Therefore, as long as average multiplicities are concerned, the 
calculation can be done within Boltzmann statistics and the difference between 
the correct formula and the approximate one is almost irrelevant. Thus, the only
effective requirement on cluster size for the validity of all performed analyses 
in high energy collisions \cite{beca,becapt} is that it must be larger than Compton 
wavelenght of particles (at most 1.4 fm) and this is always met. 

Even though average multiplicities are essentially unaffected in most practical
cases, there are other observables which are sensitive to quantum statistics 
effects and for which the fully correct calculation of phase space volume~\ref{clusexp}
is compelling, e.g. Bose-Einstein correlation spectra and multi-pion exclusive channel 
rates. 

\section{Microcanonical partition function}

The overall phase space volume of the ideal hadron-resonance gas is obtained 
by summing $\Omega_{\Nj}$ over all allowed channels: 
\begin{equation}\label{phspvol}
\Omega = \sum_{\Nj} \Omega_{\Nj} \delta_{\Qz,\Qs}
\end{equation}
As $\Omega_{\Nj}= \Gamma_{\Nj}/|\eta_i|^2$, $\Omega$ can be expressed on the basis
of Eq.~(\ref{projt}) after having removed the two identity resolutions in $f'$ and
$f''$:
\begin{eqnarray}\label{phspvol2}
 \Omega &=& \frac{1}{|\eta_i|^2} \sum_f \Gamma_f \nonumber \\ 
&=& \sum_{f} \sum_{h_V, h'_V} \langle f | h_V \rangle 
\langle h_V | \delta^4 (P- P_{\rm op}) \, \delta_{\Qz,\Qz_{\rm op}} | h'_V \rangle 
\langle h'_V | f \rangle \nonumber \\
&=& \sum_{h_V} \langle h_V | \delta^4 (P- P_{\rm op}) \, \delta_{\Qz,\Qz_{\rm op}}
| h_V \rangle 
\end{eqnarray}
The last expression makes it apparent that the definition of $\Omega$ as the 
{\em microcanonical partition function} is an appropriate one. If the sums in 
Eqs.~(\ref{phspvol}) and (\ref{phspvol2}) are performed over all channels regardless 
of their charge, the obtained quantity is defined as {\em grand-microcanonical 
partition function}:
\begin{equation}\label{defcorr}
 \Omega = \sum_{\Nj} \Omega_{\Nj} = \sum_{h_V} \langle h_V | \delta^4 (P- P_{\rm op}) 
 | h_V \rangle
\end{equation}
Throughout this section we will confine ourselves to the latter, rather than the to 
properly defined microcanonical partition function, in order not to bring along a 
cumbersome formalism. This limitation shall not affect the generality of the expounded 
arguments and the extension to the case of constrained charges is indeed straightforward.     

We have seen in the previous section that Eq.~(\ref{clusexp0}) is a correct
generalization of Eq.~(\ref{clusexp}) for finite volumes. Likewise, Eq.~(\ref{defcorr}) 
is a generalization of the expression quoted in previous literature \cite{hageps,becapt}:
\begin{equation}\label{defold}
\Omega = \sum_{\rm states} \delta^4(P - P_{\rm state})  
\end{equation}
In fact, Eq.~(\ref{defold}) is a straightforward consequence of Eq.~(\ref{defcorr}) 
if $|h_V\rangle$ is an eigenstate of energy-momentum. However, we have already 
emphasized that $|h_V\rangle$ is a localized state and its four-momentum is a well 
defined quantity only in the large volume limit, as has been discussed at the  end 
of Sect. 2. Thus, Eq.~(\ref{defold}) is consistent only if the cluster is sufficiently 
large, whereas the Eq.~(\ref{defcorr}) is always a well defined one.
On the other hand, a closed analytical integral expression for the correct 
(grand-)microcanonical partition function cannot be written. The best one can do is 
to decompose it as a sum over channels, as in Eq.~(\ref{phspvol}), and calculate 
numerically the $\Omega_{\Nj}$'s according to Eq.~(\ref{clusexp0}), which is a 
formidable task indeed. Conversely, Eq.~(\ref{defold}) does lead to a closed integral 
expression, which can be obtained by firstly expanding the Dirac delta in 
Eq.~(\ref{defold}) as a Fourier integral:
\begin{equation}\label{fourier} 
\delta^4(P-P_{\rm state}) = \frac{1}{(2 \pi)^4} \int \d^4 y \; \e^{\i (P-P_{\rm state}) 
\cdot y}
\end{equation}
and reexpressing Eq.~(\ref{defold}) as: 
\begin{equation}\label{omold}
\Omega = \frac{1}{(2 \pi)^4} \int \d^4 y \; \e^{\i P \cdot y} \sum_{\{n_{jh}\}}
\prod_{j,h} \e^{-\i n_{jh} p_{jh} \cdot y}
\end{equation}
The sum over states is in fact a sum over all possible occupation numbers of
each phase space cell. The calculation now proceeds by taking advantage of the 
commutability between sum and product in (\ref{omold}). However, unlike for 
fermions for which $n_{jh} = 0, 1$ only, the sum over occupation numbers does not 
converge to a finite value for bosons as $n_{jh}$ runs from $0$ to $\infty$. The 
convergence is recovered if the time component of $y$ is provided with a small 
negative imaginary part $-\i \varepsilon$. If we introduce such a term in 
Eq.~(\ref{omold}) the sums can be performed and the result is:
\begin{eqnarray}\label{omold2}
\Omega = && \lim_{\varepsilon \to 0} \frac{1}{(2 \pi)^4} \int_{-\infty-\i \varepsilon}^
{+\infty-\i \varepsilon} \!\!\!\!\!\!\!\! \d y^0 \int \d^3 {\rm y} \; \e^{\i P \cdot y} 
\nonumber \\
&& \times \exp\Big[ \sum_{j,h} \log (1 \pm \e^{-\i p_{jh} \cdot y})^{\pm 1} \Big]
\end{eqnarray}
where the upper sign applies to fermions, the lower to bosons. The integrand function
is in fact singular for $y=0$ and the shift of the integration contour in the complex 
plane provides a regularization prescription. The sum over phase space cells can be 
replaced, in the large volume limit, by an integration according to Eq.~(\ref{cell}),
so that $\Omega$ reads:
\begin{eqnarray}\label{omold3}
\!\!\!\!\!\! \Omega = && \!\!\!\lim_{\varepsilon \to 0} \frac{1}{(2 \pi)^4} 
\int_{-\infty-\i \varepsilon}^{+\infty-\i \varepsilon} \!\!\!\!\!\!\!\! 
\d y^0 \int \d^3 {\rm y} \; \e^{\i P \cdot y} 
\nonumber \\
\!\!\!\!\!\! && \!\!\!\!\!\! \times \exp \Big[ \sum_j \frac{(2J_j +1) V}{(2 \pi)^3} 
 \int \d^3 {\rm p} \; \log (1 \pm \e^{-\i p \cdot y})^{\pm 1} \Big]
\end{eqnarray}
We will prove in the remainder of this section that the closed expression for the 
grand-microcanonical partition function, Eq.~(\ref{omold3}), can be recovered without
invoking (\ref{defold}) and (\ref{cell}), but starting from the general 
expression~(\ref{clusexp0}) in at least two cases: 
\begin{enumerate}
\item{} for Boltzmann statistics; 
\item{} in a full quantum statistics treatment, by enforcing the approximation 
  Eq.~(\ref{asympt}), namely:
\begin{equation}\label{delta}
  \frac{1}{(2\pi)^3} \int_V \d^3 {\rm x} \; \e^{\i ({\bf p} - {\bf p}')\cdot 
  {\bf x}} \simeq \delta^3 ({\bf p} - {\bf p}')
\end{equation}
\end{enumerate}  
Henceforth, we will adopt the following shorthand:
\begin{equation} 
  \inty = \int_{-\infty-\i \varepsilon}^{+\infty-\i \varepsilon} \!\!\!\!\!\!\!\! 
   \d y^0 \int \d^3 {\rm y} \;
\end{equation}
1. Let us start by showing that for Boltzmann statistics. We have seen in the previous 
section that confining to classical statistics amounts to retain only the first term
${\hpartj} = (N_j, 0, \ldots, 0)$ in the general cluster decomposition Eq.~(\ref{clusexp0}), 
hence $h_1 = H_j = N_j$, $F_1 = V / (2\pi)^3$ and
\begin{eqnarray}\label{boltzmann}
  \Omega_{\Nj}^{\rm Boltz} = && \int \d^3 {\rm p}_1 \ldots \d^3 {\rm p}_N \; 
  \delta^4 (P-\sum_{i=1}^N p_i) \nonumber \\
&& \times \prod_j \frac{1}{N_j!} \Big[\frac{V (2J + 1)}{(2\pi)^3}\Big]^{N_j} 
\end{eqnarray}
The Dirac delta in the above equation can be Fourier expanded, thus, after 
regularization:
\begin{eqnarray}\label{boltzmann2} 
&& \!\!\! \Omega_{\Nj}^{\rm Boltz} = \frac{1}{(2\pi)^4} \inty \;
  \e^{\i P \cdot y} \int \d^3 {\rm p}_1 \ldots \d^3 {\rm p}_N \; \nonumber \\
&& \!\!\! \times \exp[-\i \sum_{i=1}^N p_i \cdot y] \prod_j \frac{1}{N_j!} 
  \Big[\frac{V (2J + 1)}{(2\pi)^3}\Big]^{N_j} \nonumber \\
&& \!\!\! = \frac{1}{(2\pi)^4} \intyy \; \e^{\i P \cdot y} \prod_j \frac{1}{N_j!} 
 \Big[\frac{V (2J + 1)}{(2\pi)^3} \int \d^3 {\rm p} \; \e^{-\i p_j \cdot y} \Big]^{N_j} 
\nonumber \\
&&
\end{eqnarray}
Summing over all channels yields the grand-microcanonical partition function:
\begin{eqnarray}\label{grandboltz}
\Omega^{\rm Boltz} = && \frac{1}{(2\pi)^4} \inty \; \e^{\i P \cdot y} \nonumber \\ 
&& \times \exp \Big[ \sum_j \frac{V (2J_j + 1)}{(2\pi)^3} \int \d^3 {\rm p} \; 
  \e^{-\i p_j \cdot y} \Big] 
\end{eqnarray}
which can be obtained indeed from Eq.~(\ref{omold3}) with the boltzmannian approximation:
\begin{equation}
 \log ( 1 \pm \e^{-\i p \cdot y} )^{\pm 1} \simeq \e^{-\i p \cdot y} 
\end{equation}
This proves the first part of our argument.

2. If quantum statistics is included, we make the supplementary assumption, as has 
been mentioned, that approximations (\ref{asympt}) apply and, thus, Eq.~(\ref{clusexp0}) 
turns to Eq.~(\ref{clusexp}). Let us first restore $p_{l_j} = p'_{l_j}/n_{l_j}$ 
(see Eq.~(\ref{lumping})) as integration variables and rewrite Eq.~(\ref{clusexp}) by
plugging in the Fourier expansion of the Dirac delta:
\begin{eqnarray}\label{clusexp2}
 && \Omega_{\Nj} = 
 \Bigg[ \prod_j \sum_{\hpartj} (\mp 1)^{N_j + H_j} \frac{1}{\prod_{n_j=1}^{N_j}n_j^{h_{n_j}} 
    h_{n_j}!} \nonumber \\
&& \times \prod_{l_j=1}^{H_j} \Big[\frac{V (2J_j+1)}{(2\pi)^3} 
  \int \d^3 {\rm p}_{l_j} \Big] \Bigg] \; \delta^4 (P_i - \sum_j \sum_{l_j=1}^{H_j} n_{l_j} 
   p_{l_j})  \nonumber \\ 
&& = \frac{1}{(2\pi)^4} \inty \; \e^{\i P \cdot y} \Bigg[ 
  \prod_j \sum_{\hpartj}  \frac{(\mp 1)^{N_j + H_j}}{\prod_{n_j=1}^{N_j}n_j^{h_{n_j}} 
    h_{n_j}!} \nonumber \\
&& \times \prod_{l_j=1}^{H_j} \Big[\frac{V (2J_j+1)}{(2\pi)^3} \int \d^3 {\rm p}_{l_j}\Big]
   \Bigg] \; \exp[-\i \sum_j \sum_{l_j=1}^{H_j} n_{l_j} p_{l_j} \cdot y]  \nonumber \\
&& =\frac{1}{(2\pi)^4} \inty \; \e^{\i P \cdot y} 
  \prod_j \sum_{\hpartj}  \frac{(\mp 1)^{N_j + H_j}}{\prod_{n_j=1}^{N_j}n_j^{h_{n_j}} 
    h_{n_j}!} \nonumber \\
&& \times \prod_{l_j=1}^{H_j} \Big[\frac{V (2J_j+1)}{(2\pi)^3} \int \d^3 {\rm p}_{l_j}
   \e^{-\i n_{l_j} p_{l_j} \cdot y} \Big]  
\end{eqnarray}
In order to simplify the notation, we introduce the quantities:
\begin{equation}\label{zetas}
 z_{j(n)}\equiv z_{j(n)}(y) = \frac{V (2J_j+1)}{(2\pi)^3} \int \d^3 {\rm p} \;
 \e^{-\i n p \cdot y}
\end{equation}
so that Eq.~(\ref{clusexp2}) can be further written as:
\begin{eqnarray}\label{clusexp3}
 && \Omega_{\Nj} = \frac{1}{(2\pi)^4} \inty \; 
\e^{\i P \cdot y} \nonumber \\
 && \times \prod_j \sum_{\hpartj} (\mp 1)^{N_j + H_j} 
  \frac{\prod_{l_j=1}^{H_j} z_{j(n_{l_j})}}{\prod_{n_j=1}^{N_j} n^{h_{n_j}} h_{n_j}!} 
\nonumber \\
 && = \frac{1}{(2\pi)^4} \inty \; \e^{\i P \cdot y} 
  \prod_j \sum_{\hpartj} \prod_{n_j=1}^{\infty}\!\! \frac{(\mp 1)^{(n_j+1)h_{n_j}} 
  z_{j(n_j)}^{h_{n_j}}}{n_j^{h_{n_j}} h_{n_j}!} \nonumber \\
 && 
\end{eqnarray}
where, in the last passage, we have taken advantage of the fact that $z_j(n)$ 
is constant over a conjugacy class. 
Also note that we have released the upper limit in the sum because of the constraint
$\sum_{n_j} n_j h_{n_j} = N_j$ which effectively sets it to $N_j$.
At this stage, the key observation is that we can implement this constraint through 
an integration in the complex plane for each species $j$ and then perform an 
unconstrained sum over all $h_{n_j}$:
\begin{equation}
\! \sum_{\hpartj} = \!\!\!\! \sum_{h_1,\ldots,h_{\infty}} \!\!\!\!\! 
\delta_{\Sigma_{n_j} n_j h_{n_j}, N_j} = \frac{1}{2 \pi \i} \oint 
\frac{\d w}{w^{N_j+1}} \!\! \prod_{n_j=1}^\infty \sum_{h_{n_j}=0}^\infty \!\!
 w^{n_j h_{n_j}} 
\end{equation}
so that the part of integrand in Eq.~(\ref{clusexp3}) following the $j$-product sign
can be written as:
\begin{eqnarray}\label{integrand}
&& \frac{1}{2 \pi \i} \oint \frac{\d w_j}{w_j^{N_j+1}} \prod_{n_j=1}^\infty 
 \sum_{h_{n_j}=0}^\infty \frac{(\mp 1)^{(n_j+1)h_{n_j}} z_{j(n_j)}^{h_{n_j}} 
 w^{n_j h_{n_j}}}{n_j^{h_{n_j}} h_{n_j}!} \nonumber \\
&& = \frac{1}{2 \pi \i} \oint \frac{\d w_j}{w_j^{N_j+1}} \prod_{n_j=1}^\infty 
 \exp\Big[\frac{(\mp 1)^{n_j+1}}{n_j} z_{j(n_j)} w_j^{n_j}\Big] \nonumber \\
&& = \frac{1}{N_j!} \frac{\d^{N_j}}{\d w_j^{N_j}} \exp\Big[\sum_{n_j=1}^\infty 
 \frac{(\mp 1)^{n_j+1}}{n_j} z_{j(n_j)} w_j^{n_j}\Big] \Big|_{w_j=0} 
\end{eqnarray}
By using the explicit expression of $z_{(n)}$ in Eq.~(\ref{zetas}), the series in the 
exponential of Eq.~(\ref{integrand}) can be summed up and this yields, for the phase space 
volume $\Omega_{\Nj}$:
\begin{eqnarray}\label{omegan}
\!\!\!\!  && \Omega_{\Nj}= \frac{1}{(2\pi)^4} \inty \; \e^{\i P \cdot y} 
 \prod_j \frac{1}{N_j!} \nonumber\\
\!\!\!\! && \times \frac{\d^{N_j}}{\d w_j^{N_j}} \exp\Big[\frac{(2J_j+1)V}{(2\pi)^3} 
 \!\!\! \int \d^3{\rm p}\; \log (1\pm w_j \e^{-\i p \cdot y})^{\pm 1}\Big] \Big|_{w_j=0}
\end{eqnarray}
We are now in a position to calculate $\Omega$ by summing over all $N_j \;\; \forall j$, 
according to Eq.~(\ref{defcorr}). The sum over each $N_j=0,\ldots,\infty$ can be performed 
independently and, noticing that each term is the $N_j^{\rm th}$ one of the Taylor expansion 
of the exponential function evaluated at $w_j=1$, one obtains:
\begin{eqnarray}
\!\!\!\!\!\!\!\!\! \Omega = && \frac{1}{(2\pi)^4} \inty \; \e^{\i P \cdot y} 
\nonumber \\
\!\!\!\!\!\!\!\!\!  && \times \prod_j \exp\Big[\frac{(2J_j+1)V}{(2\pi)^3} 
\int \d^3{\rm p} \; \log (1\pm \e^{-\i p \cdot y})^{\pm 1} \Big]
\end{eqnarray}
which coincides with Eq.~(\ref{omold3}); this proves our second statement.

The recovery of the known expression of the microcanonical partition function 
in the two considered cases is not surprising as the same holds for the single
channel phase space volume $\Omega_{\Nj}$. We have seen this in Sect. 2 where 
it has been emphasized that $\Omega_{\Nj}$ in Boltzmann statistics does not differ 
from its approximation in the large volume limit; and in Sect. 3, where we have 
seen that the $\Omega_{\Nj}$ in full quantum statistics (\ref{clusexp}) deduced 
from Eq.~(\ref{clusexp0}) by enforcing the approximation (\ref{asympt}), was obtained 
in the traditional approach \cite{hageps} using Eqs.~(\ref{hpspace}),(\ref{pstate}) 
and (\ref{cell}).

\section{From microcanonical to canonical ensemble}

What has been done for the grand-microcanonical ensemble can be straightforwardly
extended to the properly called microcanonical ensemble by adding the further constraint of
$M$ abelian charges conservation, like in Eq.~(\ref{phspvol}). The Kronecker delta
can be Fourier expanded:
\begin{eqnarray}\label{chcon}
 \delta_{\Qz,\Qs} &=& \prod_{m=1}^M  \frac{1}{2\pi} \int_{-\pi}^{\pi} \d \phi_m \; 
 \e^{\i (Q_m - Q_{\Nj m}) \phi_m} \nonumber \\    
 & = &  \frac{1}{(2\pi)^M} \int_{-\pivs}^{+\pivs} \d^M \phi \; \e^{\i (\Qz - \Qs) \cdot
 \phivs}
\end{eqnarray}
where the vector notation $\phiv = (\phi_1,\ldots,\phi_M)$ has been introduced. 
The reasoning in the previous section, from Eq.~(\ref{boltzmann}) onwards, can be 
easily repeated with the additional charge constraint (\ref{chcon}), under the
same conditions for the validity of the needed approximation~(\ref{delta}). 
One can thus arrive at the following expression of the microcanonical partition 
function:
\begin{eqnarray}\label{micropart}
\!\!\!\!\!\!\!\!\!\!\! && \Omega= \frac{1}{(2\pi)^{4+M}} \int \d^4 y \;\e^{\i P \cdot y} 
\int_{-\pivs}^{+\pivs} \d^M \phi \; \e^{\i \Qz \cdot \phivs} \nonumber\\
\!\!\!\!\!\!\!\!\!\!\! && \times \exp\Big[\sum_j \frac{(2J_j+1)V}{(2\pi)^3} \!\!\! 
 \int \d^3{\rm p} \; \log (1\pm \e^{-\i p \cdot y -\i \qj \cdot \phivs})^{\pm 1}\Big]  
\end{eqnarray}
where $\qj = (q_{j1},\ldots,q_{jM})$ are the abelian charges of the $j^{\rm th}$ hadron
species. Let us perform a rotation in the four-dimensional complex hyperplane by setting
$z=\i y$ and rewrite Eq.~(\ref{micropart}) as:
\begin{equation}\label{micropart2}
\Omega = \lim_{\varepsilon \to 0} \frac{1}{(2\pi\i)^4} 
\intz \;\exp[P \cdot z + \log Z(z,\Qz)]  
\end{equation}
where:
\begin{eqnarray}\label{canonical}
 \!\!\!\!\!\!\!\!\!\!\! && Z(z,\Qz) =  \frac{1}{(2\pi)^M} \int_{-\pivs}^{+\pivs} 
  \d^M \phi \; \e^{\i \Qz \cdot \phivs} \nonumber \\
 \!\!\!\!\!\!\!\!\!\!\! && \times \exp\Big[\sum_j \frac{(2J_j+1)V}{(2\pi)^3} \!\!\! 
  \int \d^3{\rm p} \; \log (1\pm \e^{- z \cdot p -\i \qj \cdot \phivs})^{\pm 1}\Big] 
\end{eqnarray}
In Eq.~(\ref{canonical}) it is recognizable the expression of the canonical partition 
function \cite{beca,becapt} calculated for a {\em complex four-temperature} $z$. 
The same expression can be obtained starting from the definition:
\begin{equation}\label{canonical2}
Z(z,\Qz) = \sum_{h_V} \langle h_V | \e^{-z \cdot P_{\rm op}}\delta_{\Qz,\Qz_{\rm op}}|
h_V \rangle
\end{equation}
and proceeding in the very same way as for the microcanonical partition function. 
Particularly, the approximations (\ref{delta}) are needed to get to Eq.~(\ref{canonical}). 

If the volume and the mass of the cluster are large, one can make an approximate 
calculation of the integral in Eq.~(\ref{micropart2}) through the saddle-point expansion.
The large-valued parameter can be either volume or mass provided that density $M/V$
is a finite value, which is indeed the case of interest in the framework of the
statistical hadronization model. The saddle-point four-vector $\beta$ is determined
by enforcing the vanishing of integrand logarithmic derivative for each component
$\mu$:
\begin{equation}\label{saddle}
\frac{\partial}{\partial z^\mu} [P \cdot z + \log Z(z,\Qz)]\Big|_{z=\beta} =
P_\mu + \frac{\partial}{\partial \beta^\mu} \log Z(\beta,\Qz) = 0
\end{equation}
We assume that the above equation has one real solution (note that $Z(z)$ is 
real for real argument,see Eq.~(\ref{canonical2})). This must be a 
timelike four-vector for the momentum integration in Eq.~(\ref{canonical}) to converge. 
Therefore, we can set $\beta = (1/T) \hat u$ where $\hat u$ is a unit timelike vector 
and $T > 0$ is defined as temperature, while $\beta$ is usually called temperature 
four-vector. It is not difficult to verify that if the cluster's rest frame is chosen, 
where $P = (M,{\bf 0})$, $\beta$ has vanishing spacial components and the usual expression 
of the canonical partition function is recovered:
\begin{equation}\label{canonical3}
Z(\Qz) = \sum_{h_V} \langle h_V | \e^{- H_{\rm op}/ T}\delta_{\Qz,\Qz_{\rm op}}|
h_V \rangle
\end{equation}
where $H_{\rm op}$ is the hamiltonian. Retaining only the leading term of the 
asymptotic expansion,the microcanonical partition function can be approximated as:
\begin{equation}\label{asymexp}
  \Omega \simeq \exp[P \cdot \beta + \log Z(\beta,\Qz)] \sqrt{\frac{1}{(2\pi)^4 
\det {\sf H}(\beta,\Qz)}}
\end{equation}
where $\sf H$ is the Hessian matrix $\partial^2 \log Z/\partial z^\mu \partial z^\nu$.
In the cluster's rest frame $\beta = (\bar\beta, {\bf 0})$ as already pointed out, 
thus, according to Eq.~(\ref{canonical}), the derivative $\partial \log Z /\partial z^i$ 
with respect to the spacial components of $z$ vanish because of odd-symmetric momentum 
integrands and, consequently, the Hessian determinant in Eq.~(\ref{asymexp}) simply 
becomes $\partial^2 \log Z/ \partial \bar\beta^2 = C_V T^2$. Altogether, 
if $V$ is large, the microcanonical partition function $\Omega$ is proportional to 
the canonical partition function $Z$ and we can write:
\begin{equation}\label{approx}
\Omega (P,\Qz) \underset{V \to \infty}{\propto} e^{\beta \cdot P} Z(\beta, \Qz) 
\end{equation}
with $\beta$ given by Eq.~(\ref{saddle}), and:
\begin{equation}
Z(\beta, \Qz) = \int \d^4 P \; \theta(P^0) \, \e^{-\beta \cdot P} \Omega (P,\Qz)
\end{equation}
This equation is indeed an exact one, as can be realized from Eq.~(\ref{micropart}); 
the canonical partition function is in fact the Laplace transform of the microcanonical one. 

The question arises whether and in which range of values of cluster's volume and 
mass the approximation~(\ref{approx}), i.e. the use of the canonical ensemble, 
employed in several analyses of multiplicities in elementary collisions, is 
a good one for the calculation of relevant physical quantities. This issue can be 
tackled only numerically for the particular system of the ideal hadron-resonance gas, 
comparing the exact with the approximate calculation; as has been mentioned in the
Introduction, this will be the main subject of the second paper \cite{micro2}.   

The way temperature has been introduced starting from the microcanonical ensemble
in Eq.~(\ref{saddle}) is rather unusual and deserves some discussion. Through the 
saddle-point relation~(\ref{saddle}), we have defined a temperature by enforcing the 
known values of energy and momentum of the cluster to be what it can be easily 
recognized as the average energy and momentum in the canonical ensemble, that is, 
in the cluster's rest frame where $P=(M, {\bf 0})$ and $\beta= (\bar\beta, {\bf 0})$:
\begin{equation}\label{saddle2}
 M  =  - \frac{\partial}{\partial \bar\beta} \log Z (\bar\beta,\Qz)
\end{equation}
with $T=1/\bar\beta$. 
On the other hand, it is also possible \cite{gross} to extend the relation:
\begin{equation}
\frac{1}{T} \equiv \frac{\partial S}{\partial M}   \qquad {\rm with} \qquad 
 S = \log \Omega
\end{equation}
to the microcanonical regime. This definition gives rise to the following equation, 
by using Eq.~(\ref{micropart2}):
\begin{equation}
\frac{1}{T} = \frac{1}{\Omega} \lim_{\varepsilon \to 0} \frac{1}{(2\pi\i)^4} 
\intz \; z^0 \e^{M  z^0} Z(z,\Qz)  
\end{equation}
which implies a different definition of temperature with respect to Eq.~(\ref{saddle2}).
At the leading order of the asymptotic expansion of the above integral and $\Omega$, 
the previous equation reads:
\begin{equation}
\frac{1}{T} \simeq \frac{\bar\beta' \exp[M + \frac{\partial}{\partial \bar\beta'} 
\log Z (\bar\beta',\Qz)](C_V(\bar\beta)\bar\beta^2)^{-1/2}}
{\exp[M + \frac{\partial}{\partial \bar\beta} \log Z (\bar\beta,\Qz)]
(C_V(\bar\beta')\bar\beta'^2)^{-1/2}}
\end{equation}
where $\bar\beta$ is the solution of Eq.~(\ref{saddle2}) and $\bar\beta'$ that of
\begin{equation}\label{saddle3}
 \frac{1}{\bar\beta'} + M  =  - \frac{\partial}{\partial \bar\beta'} 
\log Z (\bar\beta',\Qz)
\end{equation}
If the system is very large, i.e. in the thermodynamical limit, the temperature
$1/\bar\beta'$ is much less than $M$ so, according to Eq.~(\ref{saddle3}) 
$\bar\beta' \simeq \bar\beta$, $1/T \simeq \bar\beta' \simeq \bar\beta$ and the 
two definitions coincide, as expected.
 
It is worth pointing out that, even in the canonical ensemble, for finite volumes, 
$\log Z$ is not a linear function of $V$ (see Eq.~(\ref{canonical})) and this has
the remarkable consequence that $T$, in both definitions, is not a function of 
$M/V$ but of $M$ and $V$ separately. Otherwise stated, if $T$ and $V$ are used 
as independent thermodynamical parameters, the mean energy is not an extensive 
variable as it does not scale linearly with $V$. Of course, this mostly unfamiliar 
feature disappears in the thermodynamic limit.

\section{Physical observables}

The comparison of the model predictions with experimental measurements 
involves the calculation of quantities which can be always written as averages or 
expectation values of some operator. For instance, the average multiplicity of
the $j^{\rm th}$ hadron species in the grand-microcanonical ensemble can be written 
as:
\begin{equation}\label{mult}
 \langle \hat{N_j} \rangle = \frac{\sum_{\Nj} N_j \Omega_{\Nj}}{\Omega}
\end{equation}
the correlations between $j^{\rm th}$ and $k^{\rm th}$ hadron species as the 
expectation value of $ (\hat{N_j} -\langle N_j \rangle)({\hat N_k} -\langle N_j 
\rangle)$ and the probability of a single configuration $\Nj$ as the expectation value
of $\delta_{{\hat N_1},N_1},\ldots,\delta_{{\hat N_K},N_K}$. The analytical expressions of
sums like that in Eq.~(\ref{mult}) can be obtained by multiplying $\Omega_{\Nj}$
by a factor (a fictitious fugacity) $\lambda_j$ powered to $N_j$ and taking the
derivative with respect to $\lambda_j$ for $\lambda_j=1$. Therefore, for the average 
multiplicity of the $j^{\rm th}$ hadron species:  
\begin{equation}\label{mult2}
 \langle \hat N_j \rangle = \frac{\sum_{\Nj} N_j \Omega_{\Nj}}{\Omega} =
 \frac{\partial}{\partial \lambda_j} \log \sum_{\Nj} \lambda_j^{N_j} \Omega_{\Nj}
 \Big|_{\lambda_j = 1}
\end{equation}
The sum on the right hand side can be generalized to all species:
\begin{equation}\label{generat}
 G(\lambda_1,\ldots,\lambda_K) = \sum_{\Nj} \Omega_{\Nj} \prod_j \lambda_j^{N_j}
\end{equation}
and $G$ can be properly defined as the generating function of the multiparticle 
multiplicity distribution. Note that $G(1) = \Omega$. 

The main advantage of this method of expressing expectation values is that the 
generating function can be calculated analytically. By using the expression of 
$\Omega_{\Nj}$ in Eq.~(\ref{omegan}), the right hand side of Eq.~(\ref{generat}) can 
be turned into:
\begin{eqnarray}\label{ggrandm}
\!\!\!\!\!\!\!\! && G (\lambda_1,\ldots,\lambda_K) = \frac{1}{(2\pi)^4} \inty \; 
\e^{\i P \cdot y} \nonumber \\ 
\!\!\!\!\!\!\!\!  && \times \exp\Big[\sum_j \frac{(2J_j+1)V}{(2\pi)^3} \int \d^3{\rm p} \; 
\log (1\pm \lambda_j \e^{-\i p \cdot y})^{\pm 1} \Big]
\end{eqnarray}
and similarly in the canonical case. Now the expectation value of any operator 
can be calculated from the generating function by applying many times the 
differential operators $D_j = \lambda_j \partial/\partial \lambda_j$. In fact,
according to Eq.~(\ref{generat}), for the $M^{\rm th}$ power of $N_j$: 
\begin{equation}\label{expectm}
 \langle \hat N^M_j \rangle = \frac{1}{\Omega} \Big[ \prod_{i=1}^M \lambda_j 
 \frac{\partial}{\partial \lambda_j} \Big] G(\lambda_1,\ldots,\lambda_K) \Big|_{\lambda=1}
\end{equation}
Then, since the operators $D_j$ and $D_k$ commute, we can write formally, for
any function of $N_1,\ldots,N_K$:
\begin{equation}
\langle F(\hat\Nj) \rangle = \frac{1}{\Omega} F(D_1,\ldots,D_K) 
   G(\lambda_1,\ldots,\lambda_K)
 \Big|_{\lambda=1}
\end{equation} 
Thereby, analytical expressions of various observables can be inferred. For instance, 
by using Eq.~(\ref{expectm}) with $G$ given by (\ref{ggrandm}), the average 
multiplicity of the $j^{\rm th}$ hadron, in the Boltzmann statistics limit, turns 
out to be:
\begin{equation}\label{multan}
 \langle \hat{N_j} \rangle = \frac{1}{\Omega(P)} \frac{(2J_j+1) V}{(2\pi)^3} 
 \int \d^3 {\rm p} \; \Omega (P-p_j)
\end{equation}
where $\Omega(P)$ is given by Eq.~(\ref{omold3}). It is worth remarking that,
since $\Omega(P-p_j)$ vanishes when $(P-p_j)^2 < 0$, the integration in momentum 
is cut off when, in the cluster's rest frame, the energy of the particle exceeds 
the cluster's mass, as it should naturally occur in a microcanonical framework. 

Despite their simple appearance, expressions like (\ref{multan}) are extremely hard 
to calculate analytically. In fact, the whole issue of providing closed formulae of 
multiplicities, correlations etc. reduces to the calculation of the generating 
function in Eq.~(\ref{ggrandm}). However, an explicit solution of that four-dimensional 
integral is known only in the two limiting cases of ultrarelativistic (vanishing masses) 
and non-relativistic gas \cite{lepore}. For the relativistic gas with massive 
particles, which pertains to the hadronic system, no closed formula useful
for numerical evaluation has ever been obtained, not even as a series. 
Therefore, the only practicable way of calculating averages within the 
microcanonical ensemble is to evaluate $\Omega_{\Nj}$ integral expressions
like (\ref{clusexp}) (which is in turn made up of integral terms 
like (\ref{boltzmann})) and sum over all possible channels. However, also
those integrals have been solved analytically only in the aforementioned
two limiting cases because the functions to be dealt with are essentially 
the same. Several authors have tried approximations \cite{approx} but in 
most cases it is difficult to keep the error under control so that, at 
some fixed order truncation of the expansions, the relative accuracy may vary 
from some percent to a factor of 10 \cite{cerhag1}. Thus, the problem 
of exploring hadronic microcanonical ensemble can be attacked only 
numerically through Monte-Carlo integration. This has been done by 
Werner and Aichelin in a quite recent paper with a method based on the
Metropolis algorithm \cite{weai}. In the next paper, we will present a full 
numerical calculation for the ideal hadron-resonance gas which exploits a 
modification of that method, very effective for large clusters, taking advantage of 
the grand-canonical limit of the multiplicity distributions as proposal 
matrix in the Metropolis algorithm.

\section{Summary and outlook}

This paper is the first of a series of two devoted to the study of the microcanonical 
ensemble of the hadron gas, which is the most fundamental framework for the 
statistical hadronization model. In this work we have mainly developed the 
analytical formalism, while numerical calculations will be the main subject of 
the second paper. The main achievements can be summarized as follows:
\begin{enumerate} 
\item{} We have provided a consistent formulation of the statistical hadronization 
model starting from purposely defined quantum transition probabilities. This
formulation is much easier to handle than previous ones based on time-reversal
arguments and $S$-matrix elements averaging and allows to calculate any 
final-state observables more straightforwardly. Furthermore, it is easier to
extend it to the case of angular momentum and parity conservation, whenever 
needed. We think that this formulation clarifies once more that it is possible
to account for the observed statistical equilibrium of the final state 
hadronic multiplicities as a result of prehadronic cluster decays, without 
invoking a thermalization process driven by collisions between formed hadrons. 
\item{} We have worked out the rates of exclusive channels neglecting angular 
momentum, parity, isospin and C-parity conservation (which are important only 
for very small hadronizing systems) and recovered known expressions in the 
statistical model. We have obtained an expression of the phase space volume 
in full quantum statistics as a cluster decomposition, Eq.~(\ref{clusexp0}), 
generalizing previous ones \cite{hageps} which are valid only asymptotically, 
i.e. in the limit of a very large cluster (in practice with a linear size 
roughly larger than 3-10 fm). This expression is valid provided that 
relativistic quantum field effects are neglected, i.e. the hadronizing cluster 
should be sufficiently larger than Compton wavelenghts of the hadrons. 
\item{} We have shown analytically how the canonical ensemble can be obtained 
as an approximation of the microcanonical ensemble for large volumes and mass 
of the cluster. 
\end{enumerate} 
In the second forthcoming paper \cite{micro2}, the numerical integration of the 
microcanonical expressions obtained in this paper will be carried out by means of 
a Monte-Carlo method. This will enable a detailed comparison with the canonical 
ensemble and to establish the range of validity of the latter, which has been 
used in the actual comparisons with measured hadronic multiplicities 
\cite{beca,becapt}. Besides, the implementation of a reasonably fast and 
reliable Monte-Carlo algorithm for the microcanonical hadronization of single 
clusters in high energy collisions is a decisive step for further tests of the
statistical hadronization model.

\section*{Acknowledgements}

We are grateful to J. Aichelin and K. Werner for useful discussions.
This work has been carried out within the INFN research project FI31.

\section*{Appendix}    
\appendix
\renewcommand{\theequation}{\thesection.\arabic{equation}}

\setcounter{equation}{0}
\section{Symmetries of operator $W$}    

We briefly discuss the requirements on the operator $W$ in (\ref{trans}) for
the fulfillement of known strong interactions symmetries. If $U(g)$ is the
unitary representation onto Hilbert space of an element $g$ belonging to a 
symmetry group:
\begin{equation}
 U(g) W U(g)^{-1} = W 
\end{equation}
Thus, as $\hat\eta$ depends, by definition, only on Casimir operators:
\begin{eqnarray}
&& U(g) P_V U(g)^{-1} = P_V \nonumber \\
&& \sum_{h_V} | U(g) h_V \rangle \langle U(g) h_V | = 
\sum_{h_V} | h_V \rangle \langle h_V |
\end{eqnarray}
Since $U(g)$ is a one-to-one correspondance, the requirement is met if, for 
any $g$, every $| U(g) h_V \rangle$ is a multihadronic state of the cluster, i.e. a 
$| h'_V \rangle$ or a linear combination of them. This is obvious if $| h_V \rangle$ 
are eigenvectors of $U(g)$, which is the case for the U(1) groups associated with
abelian additive charges, and quite straightforward for isospin SU(2) since $P_V$ 
is the projector identity as far as the isospin degrees of freedom are concerned; 
also charge conjugation symmetry is trivially satisfied.

The situation is rather different for space-time symmetries. In this case, the 
translation, rotation and reflection operators transform the projector $P_V$ 
in the projector onto the translated, rotated or reflected cluster respectively; 
only if this object is the same as the starting one, symmetry is fulfilled. 
Therefore, angular momentum and parity are conserved only if the cluster is 
spherical in shape, while energy and momentum are not conserved because of
the finite volume.

\setcounter{equation}{0}
\section{Decomposition of the Poincar\'e group projector}    

The general transformation of the extended Poincar\'e group $g_z$ may be factorized 
as:
\begin{equation}
  g_z = {\sf T}(x) {\sf Z} {\sf \Lambda} = {\sf T}(x) {\sf Z} 
  {\sf L}_{\hat{\bf n}}(\xi) {\sf R}
\end{equation}
where ${\sf T}(x)$ is a translation by the four-vector $x$, ${\sf Z} = {\sf I, \Pi}$
is either the identity or the space inversion and ${\sf \Lambda} = 
{\sf L}_{\hat{\bf n}}(\xi) {\sf R}$ is a general orthocronous Lorentz transformation 
written as the product of a boost of hyperbolic angle $\xi$ along the space-like 
axis $\hat{\bf n}$ and a rotation $\sf R$ depending on three Euler angles. Thus
Eq (\ref{proj}) becomes:
\begin{eqnarray}
 &&\!\!\!\!\!\!\! {\sf P}_{P,J,\lambda,\pi} = \nonumber \\
 &&\!\!\!\!\!\!\! \frac{1}{2} \sum_{{\sf Z}={\sf I, \Pi}} 
 \frac{\dim \nu}{(2\pi)^4} \int \d^4 x \int \d {\sf \Lambda} \; D^\nu({\sf T}(x)
 {\sf Z}{\sf \Lambda})^{i*}_i \, U({\sf T}(x) {\sf Z} {\sf \Lambda}) \nonumber \\
 &&\!\!\!\!\!\!\! = \frac{1}{2} \sum_{{\sf Z}={\sf I, \Pi}} 
  \frac{\dim \nu}{(2\pi)^4} \int \d^4 x \int \d {\sf \Lambda} \; 
 \e^{\i P \cdot x} \pi^{z} D^\nu({\sf \Lambda})^{i*}_i \nonumber \\
 && \; \; \times U({\sf T}(x)) \, U({\sf Z}) \, U({\sf \Lambda}) 
\end{eqnarray} 
where $z=0$ if ${\sf Z}= {\sf I}$ and $z=1$ if ${\sf Z}= {\sf \Pi}$.
In the above equation, by $\d {\sf \Lambda}$ we meant the invariant normalized 
measure of the Lorentz group, which can be written as \cite{lgroup}:
\begin{equation}\label{measure}
  \d {\sf \Lambda} = \d {\sf L}_{\bf n}(\xi) \, \d {\sf R} =
  \sinh^2 \xi \d \xi \, \frac{\d \Omega_{\hat{\bf n}}}{4 \pi} \, \d {\sf R}
\end{equation}
$\d {\sf R}$ being the well known invariant measure of SU(2) group.

If the initial state $| i \rangle $ has vanishing momentum, i.e. $P=(M, {\bf 0})$, 
then the Lorentz transformation $\sf \Lambda$ must not involve any non-trivial 
boost transformation with $\xi \ne 0$ for the matrix element 
$D^\nu({\sf \Lambda})^{i*}_i$ not to vanish. Therefore $\sf \Lambda$ reduces to 
the rotation ${\sf R}$ and we can write:
\begin{eqnarray}
\!\!\!\!\!\!\!\! {\sf P}_{P,J,\lambda,\pi} =&& \frac{1}{2} 
 \sum_{{\sf Z}={\sf I, \Pi}} \frac{1}{(2\pi)^4} 
 \int \d^4 x \; (2J + 1) \int \d {\sf R} \; \e^{\i P \cdot x} \pi^{z} \nonumber \\
&& \times D^J({\sf R})^{\lambda *}_\lambda \, U({\sf T}(x)) \, U({\sf Z}) \, 
  U({\sf R}) 
\end{eqnarray} 
Since $[{\sf Z}, {\sf R}] = 0$, we can move the $U({\sf Z})$ operator to the right of
$U({\sf R})$ and recast above equation as:
\begin{eqnarray}
\!\!\!\!\!\!\!\!\!\!\! {\sf P}_{P,J,\lambda,\pi} =&& \frac{1}{(2\pi)^4} 
 \int \d^4 x \; \e^{\i P \cdot x} U({\sf T}(x)) \nonumber \\
\!\!\!\!\!\!\!\!\!\!\! && \!\!\!\! \times (2J + 1) \int \d{\sf R} \; 
 D^J({\sf R})^{\lambda *}_\lambda \, U({\sf R}) \, 
 \frac{{\sf I} + \pi U({\sf \Pi})}{2}  
\end{eqnarray} 
which is the Eq.~(\ref{proj2}). 

\setcounter{equation}{0}
\section{Proof of Equations (\ref{volume}) and (\ref{bec})}    

We shall prove Eq.~(\ref{volume}) in non-relativistic quantum mechanics. It is 
assumed that $|{\bf k} \rangle$ is a complete set of states in a region $A$ with 
volume $V$, with eigenfunctions $\psi_{\bf k} ({\bf r})$ and that the transformation 
from $\sigma$ to $\tau$ polarization states is unitary. Thus:  
\begin{equation}
  \langle {\bf r} \sigma | {\bf k} \tau \rangle = \left\{ 
   \begin{array}{ll} 
   \psi_{\bf k} ({\bf r}) U_{\sigma\tau} & \qquad {\rm if}  \; \; {\bf r} \in A \\ 
   0            &  \qquad {\rm if} \; \; {\bf r} \notin A 
   \end{array} \right.
\end{equation}
where $U_{\sigma\tau}$ is the element of a unitary matrix. Hence:
\begin{eqnarray}\label{volume2}
&& \sum_{{\bf k},\tau} |\langle {\bf p}\sigma | {\bf k} \tau \rangle|^2 = 
 \sum_{{\bf k},\tau} \langle {\bf p}\sigma | {\bf k} \tau \rangle 
 \langle {\bf k} \tau| {\bf p} \sigma \rangle \nonumber \\
&& = \sum_{{\bf k},\tau} \int_A \d^3 {\rm r} \int_A \d^3 {\rm r'} \;
\langle {\bf p} \sigma | {\bf r} \sigma \rangle \langle {\bf r} \sigma | 
{\bf k}\tau\rangle \langle {\bf k} \tau | {\bf r}' \sigma \rangle 
\langle {\bf r}' \sigma| {\bf p} \sigma \rangle  \nonumber \\
&& = \sum_{{\bf k},\tau} \int_A \d^3 {\rm r} \int_A \d^3 {\rm r'} \; 
 \frac{\e^{\i {\bf p} \cdot({\bf r}' - {\bf r})}}{(2\pi)^3} \, 
 \psi_{\bf k}({\bf r}) \psi_{\bf k}^*({\bf r}') |U_{\sigma\tau}|^2 
\end{eqnarray}
where we have used the normalization of the states $\langle {\bf p} | {\bf p}' \rangle 
= \delta^3 ({\bf p} - {\bf p}')$. Since the $\psi_{\bf k}$ are a complete set 
of eigenfunctions in $A$:
\begin{equation}\label{complete}
  \sum_{\bf k} \psi_{\bf k} ({\bf r}) \psi_{\bf k}^*({\bf r}') = 
  \delta^3 ({\bf r} - {\bf r}')
\end{equation}
thus, taking into account that $U$ is unitary, Eq.~(\ref{volume2}) turns to:
\begin{equation}
 \sum_{{\bf k},\tau} |\langle {\bf p}\sigma | {\bf k} \tau \rangle|^2 =  
 \frac{1}{(2\pi)^3} \int_A \d^3 {\rm r} \int_A \d^3 {\rm r'} \; 
 \delta^3 ({\bf r} - {\bf r}') = \frac{V}{(2\pi)^3}
\end{equation}
QED. 

Likewise, the Eq.~(\ref{bec}) can be proved by calculating:
\begin{eqnarray}\label{bec2}
&& \sum_{{\bf k},\tau} \langle {\bf p}_1\sigma_1 | {\bf k} \tau \rangle 
\langle {\bf k} \tau | {\bf p}_2\sigma_2 \rangle =  \nonumber \\
&& = \sum_{{\bf k},\tau} \int_A \d^3 {\rm r} \int_A \d^3 {\rm r'} \;
\langle {\bf p}_1 \sigma_1 | {\bf r} \sigma_1 \rangle \langle {\bf r} \sigma_1 | 
{\bf k}\tau\rangle \langle {\bf k} \tau | {\bf r}' \sigma_2 \rangle 
\nonumber \\
&& \langle {\bf r}' \sigma_2| {\bf p}_2 \sigma_2 \rangle  \nonumber \\
&& = \sum_{{\bf k},\tau} \int_A \d^3 {\rm r} \int_A \d^3 {\rm r'} \; \
\frac{\e^{\i {\bf p}_2 \cdot {\bf r}' - \i {\bf p}_1 \cdot {\bf r}}}{(2\pi)^3} 
\, \psi_{\bf k}({\bf r}) \psi_{\bf k}^*({\bf r}') 
U_{{\sigma_1}\tau} U_{{\sigma_2}\tau}^* \nonumber \\
&& = \delta_{{\sigma_1},{\sigma_2}} \frac{1}{(2\pi)^3} 
\int_A \d^3 {\rm r} \; \e^{\i {\bf r} \cdot ({\bf p}_2 - {\bf p}_1)}
\end{eqnarray}
where the Eq.~(\ref{complete}) and unitarity of $U$ have been used.

\setcounter{equation}{0}
\section{Extra strangeness suppression}    

The use of an extra strangeness suppression parameter $\gs$ is quite common
in statistical model analyses in canonical and grand-canonical ensembles.
We show here how this parameter can be inserted in the microcanonical 
ensemble giving rise to the usual formulae in the large volume limit. All 
that is needed is to multiply $W$ by an operator which add a factor $\gs$ 
for each {\em pair} of valence strange quarks which is created or destroyed 
in the final state. Thus Eq.~(\ref{prob}) becomes: 
\begin{equation}
 \Gamma_f \rightarrow \Gamma_f \gs^{|N_{Sf}-N_{Si}|}
\end{equation}
where $N_{Si}$ the number of strange quarks in the initial state and 
$N_{Sf} = \sum_j N_j s_j$ that in the final state, $s_j$ being the number of
valence strange quarks in the $j^{\rm th}$ hadron species. Then, it is quite 
straightforward to extend the formulae shown in this paper for the presence 
of this additional factor. 
In particular, if $N_{Si}=0$, the rate can be written:
\begin{equation}
 \Gamma_f = |\eta_i|^2 \Omega'_{\Nj}
\end{equation}
where
\begin{equation}\label{omega'}
 \Omega'_{\Nj} = \gs^{\sum_j N_j s_j} \Omega_{\Nj}
\end{equation}
and $\Omega_{\Nj}$ as quoted throughout the paper. As far as mesons with fractional 
content $C_S \in [0,1]$ of $\ssb$ are concerned ($\eta$ for instance), an independent
incoherent superposition of the rates is assumed as though the meson was a $\ssb$ 
state in a fraction $C_S$ of observed reactions. Therefore Eq.~(\ref{omega'}) must 
be rewritten as:
\begin{equation}
 \Omega'_{\Nj} = \prod_j f_j^{N_j} \Omega_{\Nj}
\end{equation}
where:
\begin{equation}
  f_j = \left\{ 
   \begin{array}{ll} 
   1-C_{Sj} + C_{Sj} \gs^2  &  \qquad {\rm for\; unflavoured \; mesons} \\ 
   \gs^{s_j}                &  \qquad {\rm otherwise}  
   \end{array} \right.
\end{equation}
These modifications lead to a corresponding modification of the microcanonical 
partition function:
\begin{equation}
\Omega' = \sum_{\Nj} \Omega'_{\Nj} =  \sum_{\Nj} f_j^{N_j} \Omega_{\Nj}
\end{equation}
Under the same condition of validity of the approximations, it can be proved, 
going along the equations quoted in Sect.~4, that this expression can be calculated 
explicitely. The grand-microcanonical partition function reads:
\begin{eqnarray}
 && \!\!\!\!\!\!\!\!\!\!\!\! \Omega= \frac{1}{(2\pi)^4} \inty \; 
  \e^{\i P \cdot y} \nonumber \\
 && \!\!\!\!\!\!\!\!\!\!\!\! \times \exp\Big[ \sum_j \frac{(2J_j+1)V}{(2\pi)^3} 
 \int \d^3{\rm p} \; \log (1\pm f_j \e^{-\i p \cdot y})^{\pm 1} \Big]
\end{eqnarray}

The canonical partition function can be obtained starting from the
above microcanonical partition function like in Sect.~5:
\begin{eqnarray}
&& Z(\beta,\Qz) = \frac{1}{(2\pi)^M} \int_{-\pivs}^{+\pivs} \d^M \phi \; 
\e^{\i \Qz \cdot \phivs} \nonumber \\
&& \times \exp\Big[\sum_j \frac{(2J_j+1)V}{(2\pi)^3} \int \d^3{\rm p} 
\; \log (1\pm f_j \e^{- \beta \cdot p -\i \qj \cdot \phivs})^{\pm 1}\Big] 
\nonumber \\
&&
\end{eqnarray}
which is the same as usually employed to derive the hadron multiplicities 
\cite{beca}.


\end{document}